\documentclass{aa} 

\usepackage{graphicx}
\usepackage{txfonts}
\usepackage{CJK}
\usepackage{natbib}
\bibpunct{(}{)}{;}{a}{}{,}
\usepackage{hyperref}
\hypersetup{
    colorlinks=true,
    linkcolor=blue,
    filecolor=blue,      
    urlcolor=blue,
    citecolor=blue
}
\usepackage{booktabs}
\usepackage{threeparttable}
\usepackage{makecell}

\newcommand{\EBV}{E_{\rm B{-}V}}
\newcommand{\Av}{A_{\rm V}}
\newcommand{\teff}{T_{\rm eff}}
\newcommand{\logg}{\log{g}}
\newcommand{\meta}{{\rm [M/H]}}

\newcommand{\kms}{\rm km\,s^{-1}}
\newcommand{\Wdib}{W_{\rm DIB}}
\newcommand{\Wgaia}{W_{8621}}
\newcommand{\Rv}{R_{\rm V}}
\newcommand{\EBPRP}{E(G_{\rm BP}\,{-}\,G_{\rm RP})}
\newcommand{\BPRP}{(G_{\rm BP}\,{-}\,G_{\rm RP})}

\newcommand{\SEx}{{\rm log_{10}}(A_{\rm V})}
\newcommand{\SEy}{{\rm log_{10}}(W_{8621}/A_{\rm V})}
\newcommand{\UVflux}{\Phi_{\rm UV}}

\defcitealias{Sonnentrucker1997}{S97}
\defcitealias{Lallement2022}{L22}
\defcitealias{Edenhofer2024}{E24}

\makeatletter
\renewcommand*\aa@pageof{, page \thepage{} of \pageref*{LastPage}}
\makeatother

\begin{document}

\begin{CJK*}{UTF8}{gbsn}

\title{A cationic carrier for diffuse interstellar band at 862.1 nm: Evidence from the skin effect in 
nearby diffuse-to-translucent clouds}

\author{H. Zhao (赵赫) \inst{1,2} \thanks{Corresponding author: He Zhao}
        \and
        L. Li (李璐) \inst{3}}

\institute{Chinese Academy of Sciences South America Center for Astronomy (CASSACA), Camino El Observatorio 1515, 
           Las Condes, Santiago Chile \\
           \email{he.zhao@oca.eu} 
           \and 
           Departamento de Fisica y Astronomia, Facultad de Ciencias Exactas, Universidad Andres Bello, 
           Fernandez Concha 700, 8320000 Santiago, Chile 
           \and
           Shanghai Astronomical Observatory, Chinese Academy of Sciences, 80 Nandan Road, Shanghai 200030, China
          }




\abstract
{The tendency of some diffuse interstellar band (DIB) carriers to concentrate in the outer, UV-illuminated layers of molecular 
clouds (MCs)--the ``skin effect''--makes their spatial distribution a powerful probe of their physical nature. We leverage Gaia 
DR3 measurements of the DIB at 862.1\,nm to investigate its behavior across 12 nearby MCs, spanning diffuse to translucent regimes 
($\Av\,{\sim}\,0.2{-}3.5$\,mag). We find significant diversity in the DIB behavior, both between different clouds and within 
individual clouds from their outer to inner regions. To quantify these trends, we employed a piecewise linear model (PLM) to fit 
the average slope ($\alpha$) between the normalized DIB strength, $\SEy$, and dust extinction, $\SEx$. In general, $\SEy$ declines 
with $\SEx$ with $\alpha$ between 0 and --1, becoming progressively steeper at higher $\Av$. These observed slopes and their 
variations are consistent with the photoionization equilibrium models, where the carrier abundance is governed by local conditions, 
particularly the UV radiation field and cloud structure (e.g., density profiles, clumpiness). Particularly, the Taurus cloud region
uniquely displays an initial increase in $\SEy$ at low extinction, a signature predicted for a cationic carrier. By fitting the slope 
of this rising trend, we estimate an ionization potential of $E_{\rm IP}\,{=}\,12.40^{+1.90}_{-2.29}$\,eV 
for the DIB\,$\lambda$8621 carrier, which aligns well with 
the secondary ionization energies of large carbonaceous molecules like polycyclic aromatic hydrocarbons (PAHs) or fullerenes. 
Furthermore, by comparing the peak position of ${\rm log_{10}}(\Wgaia/\EBV$) in Taurus with other prominent optical DIBs, we propose 
a spatial sequence within clouds, placing the $\lambda$8621 carrier's layer between those of $\lambda$5780 and $\lambda$6614. 
Our results support a cationic carrier for DIB\,$\lambda$8621 and help situate it within the broader context of the DIB family.}







\titlerunning{Skin effect}
\maketitle

\section{Introduction} \label{sect:intro}

Diffuse interstellar bands (DIBs) are absorption features ubiquitously observed in the interstellar medium (ISM) of the Milky Way 
\citep{Kos2014, Zasowski2015c, Schultheis2023DR3}, external galaxies \citep{Cordiner2011, Monreal-Ibero2018, Erp2025}, and 
distant quasars \citep{Junkkarinen2004, York2006, Lawton2008}. 
Over 600 DIBs have been confirmed between 0.4 and 2.4\,$\rm \mu m$ 
\citep{Hobbs2008,Hobbs2009,Cox2014,Fan2019,Hamano2022,Ebenbichler2022}.
Despite their ubiquity, the origin and nature of the carriers responsible for these weak and broad absorption features 
have remained a mystery for over a century since their discovery \citep{Heger1922}. 
Currently, complex carbon-bearing molecules are widely considered the most promising candidates for DIB carriers. This view is 
supported by the successful laboratory identification of $C_{60}^+$ as the carrier of five DIBs between 9300 and 9700\,{\AA} 
\citep{Campbell2015, Walker2016, Linnartz2020} and by the detection of resolved substructures in the profiles of 
DIBs\,$\lambda$5797\footnote{We cite DIBs by their central rest-frame wavelengths in angstroms.}, $\lambda$6379, and $\lambda$6614 
in high-resolution spectra. These substructures provide evidence for unresolved rotational contours arising from the electronic 
transitions of large molecules \citep{Sarre1995, Huang2015, MacIsaac2022}. Numerous candidates have been proposed, including 
carbon chains \citep{Maier2004}, polycyclic aromatic hydrocarbons \citep[PAHs;][]{Salama1996, EF1996, Sonnentrucker1997, Ruiterkamp2005}, 
fullerenes \citep{Webster1993, Fulara1993b, Omont2016}, and polyacenes \citep{Omont2019}, but none have yet been definitively confirmed.

In addition to laboratory measurements and high-resolution spectroscopy, studying the behavior of DIB strengths in different 
environments relative to known ISM species is another key method for inferring the properties of DIB carriers and constraining 
their nature. A linear relationship between dust extinction and the equivalent width ($\Wdib$) of a DIB has been widely 
reported since the 1930s \citep[e.g.,][]{MW1938, Friedman2011, Lan2015}. However, this correlation is often attributed to the 
general increase in the abundance of all ISM species with distance along a sightline and does not always hold \citep{Krelowski2018}. 
Indeed, observations have revealed different vertical scale heights for dust grains and DIB carriers \citep{hz2023, Lallement2024}. 
For sightlines probing isolated cloud regions, $\Wdib$ is often weaker than expected from the dust extinction and eventually 
stops to grow with increasing cloud opacity, which implies that the majority of DIB carriers are concentrated in the outer layers 
of clouds \citep{Herbig1995, Snow2014iaus}. This phenomenon is commonly referred to as the ``skin effect'', first reported by 
\citet{Wampler1966}. \citet{SC1974} systematically studied the skin effect for DIBs $\lambda$4428 and $\lambda$5780 in the $\rho$
Ophiuchus cloud, followed by reports for other strong optical and near-infrared DIBs 
\citep[e.g.,][]{MU1984, Adamson1991, Adamson1994, Sonnentrucker1997, Snow2002b, Vos2011, EL2019}.

The skin effect is an environmental phenomenon, reflecting changes in the life cycle and charge state of DIB carriers as they move 
from the outer to the inner regions of clouds under varying physical conditions, such as the local ultraviolet (UV) radiation field, 
temperature, and density. By combining observations of the Orion {\ion{H}{ii}} and Taurus--Perseus regions,
\citet[][hereafter \citetalias{Sonnentrucker1997}]{Sonnentrucker1997}
observed a two-stage behavior for four optical DIBs: $\lambda$5780, $\lambda$5797, 
$\lambda$6379, and $\lambda$6614. Their normalized strength, ${\rm log_{10}}(\Wdib/\EBV)$, first increased up to $\EBV\,{\sim}\,$0.2--0.3\,mag
and then began to decrease out to $\EBV\,{\sim}\,1$\,mag. This two-stage behavior was also noted by \citet{Jenniskens1994} for 
$\lambda$6284 in Orion region. \citetalias{Sonnentrucker1997} developed a photoionization equilibrium (PIE) model to infer the 
ionization properties of DIB carriers, assuming a constant electron density and uniform cloud opacity. In their model, the 
initial increase in ${\rm log_{10}}(\Wdib/\EBV)$ occurs in UV-penetrated regions where the DIB carrier abundance is inversely proportional 
to the UV flux ($\UVflux$). The positive slope of the ${\rm log_{10}}(\Wdib/\EBV)$ versus $\EBV$ relation is determined by the dust extinction 
curve ($\Rv$) and the ionization potential ($E_{\rm IP}$) of the carrier. Deeper within the cloud, at higher $\EBV$, the carrier abundance 
decreases as $\UVflux$ drops exponentially with optical depth. The carrier abundance reaches a limit ($\Wdib$ becomes constant) once 
the UV radiation is fully attenuated. This scenario predicts that DIB carriers are concentrated in a specific layer of a cloud and is 
supported by the observed sensitivity of DIB strengths and profiles to the UV radiation field \citep{Jenniskens1994, Cami1997, Vos2011}. 
Based on their estimated $E_{\rm IP}$ of around 10\,eV, \citealt{Sonnentrucker1997} proposed ionized PAHs as carriers for DIBs 
$\lambda$5780, $\lambda$5797, and $\lambda$6614. 

Different DIBs are observed to reach their maximum $\Wdib/\EBV$ at different $\EBV$ values (\citealt{SC1974}, \citetalias{Sonnentrucker1997}, 
\citealt{Snow2002b}), and a given DIB can exhibit different behavior with respect to dust extinction in different clouds \citep{EL2019}. 
In this work, we leverage the large DIB\,$\lambda$8621 catalog from the third data release of Gaia \citep[Gaia DR3;][]{Vallenari2023, 
Schultheis2023DR3} to statistically investigate the variation of its normalized strength, $\SEy$, with respect to dust extinction, 
$\SEx$, across diffuse to translucent regions ($A_V\,{\sim}\,0.2{-}3.5$\,mag) toward 12 nearby molecular clouds (MCs). 
Our main goals are: 1) To infer the ionization properties of the $\lambda$8621 carrier using the \citetalias{Sonnentrucker1997} PIE model; 
and 2) To quantitatively characterize the dependence of the $\lambda$8621 carrier's behavior on local environmental conditions. 
This paper is organized as follows: The DIB catalog, the calculation of $\Av$, and the target selection are described in Sect. 
\ref{sect:data}. Section \ref{sect:method} introduces our model used to fit the average slopes between $\SEx$ and $\SEy$. The 
results are presented in Sect. \ref{sect:results}, and discussions are provided in Sect. \ref{sect:discuss}. Finally, we summarize 
our main conclusions in Sect. \ref{sect:conclusion}.

\section{Data and target clouds} \label{sect:data}

\subsection{DIB catalog} \label{subsect:dib_cata}

The Gaia DR3 catalog contains a vast number of DIB\,$\lambda$8621 measurements, derived from spectra of background stars observed by 
the Gaia Radial Velocity Spectrometer \citep[RVS;][]{Cropper2018, Sartoretti2018, Sartoretti2023}. These measurements were obtained 
by fitting the DIB profile in residual interstellar spectra. The interstellar spectra were produced by subtracting synthetic stellar 
templates, generated by the general stellar parameterizer from spectroscopy (GSP-Spec) module \citep{Recio-Blanco2023} within the 
astrophysical parameters inference system \citep[Apsis;][]{Creevey2023}. A detailed description of the fitting procedure, quality 
control, and validation processes is provided in \citet{hz2021a} and \citet{Schultheis2023DR3}. After removing entries with poor-quality 
DIB parameters or unreliable stellar templates, \citet{Schultheis2023DR3} defined a high-quality (HQ) sample containing approximately 
140\,000 sightlines. However, even this HQ sample has been shown to suffer from biases caused by contamination from nearby {\ion{Fe}{i}} 
stellar lines \citep{Saydjari2023, hz2024a}. Therefore, we applied a further set of filtering criteria to the HQ sample:

\begin{enumerate}
    \item The DIB quality flag needs to be zero, which indicates the highest measurement quality (\citealt{Schultheis2023DR3}).
    \item The DIB radial velocity\footnote{We applied the rest-frame wavelength of 8623.141\,{\AA} determined by \citet{hz2024a} to 
    calculate the radial velocity.}, expressed in the kinematical local standard of rest, is required to be within 
    $\pm$50\,$\kms$ as the target clouds are nearby \citep[see e.g.,][]{Wakker1991,Lewis2022}.
    \item Sightlines towards background stars with $\teff\,{=}\,4250$\,K and $\logg\,{=}\,1.5$
    are excluded to avoid the GSP-Spec parameterisation problems encountered for cool stars 
    (see Sect. 8.6 of \citealt{Recio-Blanco2023}).
    \item The Gaussian width of the DIB profile needs to be greater than 1.2\,{\AA}. This threshold is imposed to exclude very 
    narrow features that are likely to be artifacts from the imperfect subtraction of stellar lines, rather than genuine DIB 
    absorption, considering the spectral resolution of RVS spectra is about 0.75\,{\AA} at 8621\,{\AA}.
\end{enumerate}

The application of these criteria yields a clean DIB catalog containing 95\,428 sightlines. 
From this catalog, a subset of 2\,622 sightlines was selected for the present analysis 
(see Sect. \ref{subsect:target} and Table \ref{tab:plm}). 
In this subset, 96\% of the background stars have $\teff\,{>}\,4000$\,K, with a minimum 
temperature of 3719\,K.

The primary source of contamination in the measured DIB profiles arises from the residuals of stellar \ion{Fe}{i} lines. 
A precise quantification of the maximum residual for each spectrum is not possible, as the processed DIB spectra are not part 
of Gaia DR3. However, in our subsequent work \citep{Schultheis2023FPR,hz2024a}, we established that the magnitude of these stellar 
line residuals is strongly dependent on the spectral S/N. We found that for spectra with $\rm S/N\,{>}\,50$, the residuals are 
typically within 0.02, whereas they can exceed 0.05 in spectra with lower S/N.
In our sample, the median $\Wgaia$ is 0.17\,{\AA}. Given a mean Gaussian width of 2\,{\AA} for $\lambda$8621 \citep{hz2024a},
this corresponds to a central depth of 0.034 below the normalized continuum. This comparison underscores that DIBs are intrinsically 
weak features, and the use of derived ISM spectra is essential for the reliable selection of high-quality DIB measurements.

\begin{figure*}
  \centering
  \includegraphics[width=16cm]{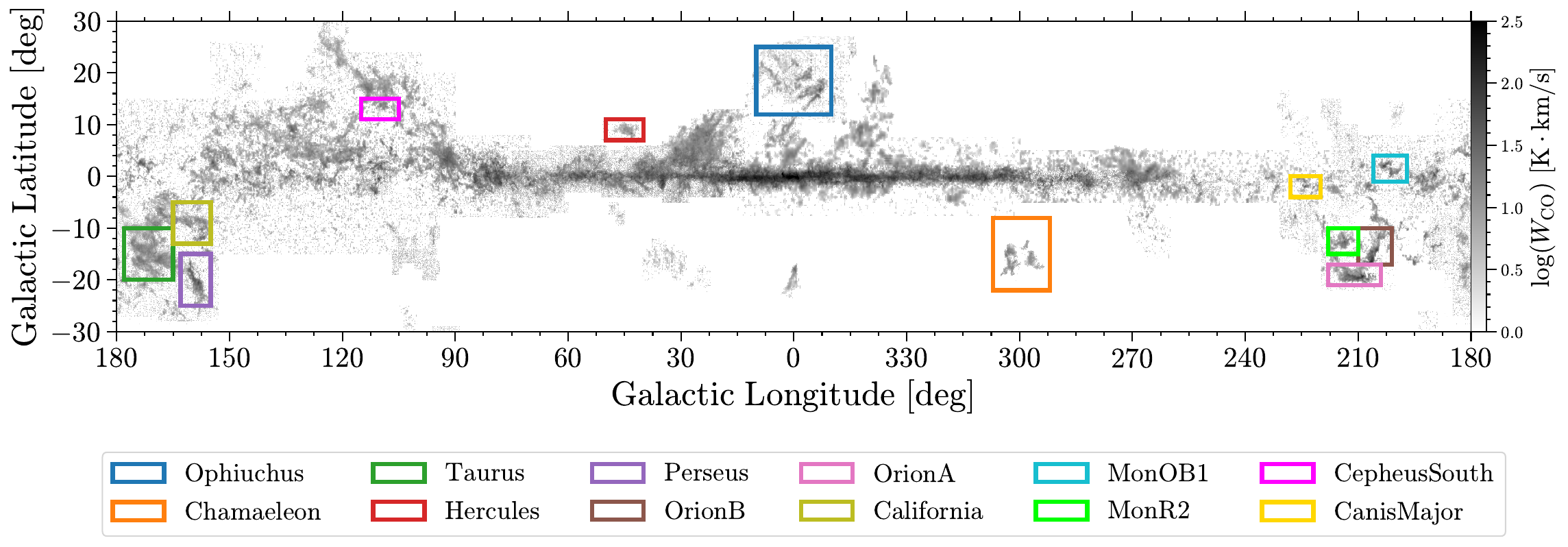}
  \caption{Location of target cloud regions (colored rectangles with names marked below), overplotted on 
  $^{12}\rm CO$ ($J\,{=}\,1{-}0$) intensity map from \citet{Dame2001}.}
  \label{fig:sky}
\end{figure*}

\subsection{Dust extinction} \label{subsect:extinction}

We determined the color excess, $\EBPRP$, for each background star by 
subtracting its predicted intrinsic color from its observed color $\BPRP$.
The intrinsic color was predicted using an XGBoost (eXtreme Gradient Boosting) model 
that takes the star's GSP-Spec atmospheric parameters--effective 
temperature ($\teff$), surface gravity ($\logg$), and metallicity ($\meta$)--as input.
XGBoost is an optimized and scalable tree-boosting library \citep{XGBoost}.
As detailed in \citet{hz2024b}, the model was trained on a sample of dust-free stars using the open-source 
package {\it xgboost}\footnote{\url{https://github.com/dmlc/xgboost}}. 
The total uncertainty on $\EBPRP$ was computed by combining three sources 
of error in quadrature: (1) the photometric errors of the star, (2) the propagated uncertainty from the atmospheric parameters, 
estimated via a Monte Carlo simulation with 1000 iterations, and (3) the intrinsic model uncertainty. To estimate the model 
uncertainty 
for a given star, we first identified its nearest neighbor within the model's test set, based on the atmospheric parameter space. 
The residual between the observed $\BPRP$ and the XGBoost prediction for that test-set neighbor was then adopted as the model 
uncertainty contribution for the target star.

To facilitate a more direct comparison with other studies and provide a more physically intuitive measure of extinction, we converted 
$\EBPRP$ to the V-band extinction, $\Av$. This conversion was performed using the standard extinction law from \citet{CCM89}, assuming 
$\Rv\,{=}\,3.1$. We did not account for potential effects from filter bandpass curvature \citep{WC2019, MaizApellaniz2024} or variations 
in $\Rv$. This simplification is justified because our analysis focuses on diffuse and translucent interstellar clouds, where extinction 
values are generally not high ($\Av\,{\lesssim}\,3.5$\,mag) and $\Rv$ is expected to not significantly vary.
A detailed investigation on the impact of $\Rv$ variation is deferred to a future paper.

\subsection{Target selection} \label{subsect:target}

In this work, we investigate the behavior of the DIB\,$\lambda$8621 in the transition from diffuse to translucent interstellar 
environments ($\Av\,{\sim}\,0.2{-}3.5$\,mag; \citealt{SM2006}). Our analysis targets 12 nearby MCs. The Galactic 
longitude/latitude ($\ell/b$) boundaries for these regions were adapted from \citet{Lee2018} and \citet{Lewis2022}, with minor 
adjustments made for this study. The cloud distances ($d_{\rm MC}$) were adopted from the references within those works. For simplicity, 
we refer to these target regions by their conventional MC names. A map showing the sky locations of these regions is presented in 
Fig. \ref{fig:sky}, and their properties are summarized in Table \ref{tab:cloud}. These specific clouds were chosen because their 
relative isolation minimizes contamination from unrelated foreground and background interstellar material 
along the line of sight.

To probe the dust extinction and DIB absorption within these clouds, we selected background stars that satisfy several criteria. 
First, a star must be located on the sky within the defined $\ell/b$ boundaries of a target cloud and have a geometric distance 
\citep{Bailer-Jones2021} between $d_{\rm MC}\,{+}\,100$\,pc and 1500\,pc, ensuring it lies behind the cloud. Second, to ensure a 
high-quality dataset, we imposed further constraints: the relative uncertainty on both the stellar distance and the DIB $\Wgaia$ 
must be less than 20\%. Finally, to ensure robust detections, we required $\Wgaia\,{>}\,0.02$\,{\AA} and $\Av\,{>}\,0.2$\,mag. 
The final number of selected sightlines for each cloud is listed in Table \ref{tab:plm}, and their spatial distributions are 
illustrated in Fig. \ref{fig:mcs}. 
This figure shows that the very densest regions of the MCs remain largely unprobed,
an observational bias resulting from high extinction combined with the limiting magnitude of the Gaia RVS instrument.
As demonstrated in Fig. \ref{fig:skin-effect}, the skin effect is evident across all target 
clouds. The median trends (blue squares) clearly reveal a diversity in the DIB-to-dust relationship, both among different clouds 
and across the diffuse-to-translucent transition within individual clouds.

\begin{table}
\centering
\tiny
\caption{Investigated cloud regions in this work sorted by their distances.}
\label{tab:cloud}
\begin{tabular}{l r@{}c@{}l r@{}c@{}l cc}
\toprule 
Name   & \multicolumn{3}{c}{Longitude range} & \multicolumn{3}{c}{Latitude range} & \multicolumn{1}{c}{Distance} & Ref.\tablefootmark{a}  \\ 
       & \multicolumn{3}{c}{(deg)}       &  \multicolumn{3}{c}{(deg)}     &  \multicolumn{1}{c}{(pc)}    &                        \\
\midrule 
Ophiuchus	  &  {$-10\,$} & {$<\ell<$} & {$\, 10$} & {$ 12\,$} & {$<b<$} & {$\, 25$} &  144 & 1 \\
Chamaeleon	  &  {$292\,$} & {$<\ell<$} & {$\,307$} & {$-22\,$} & {$<b<$} & {$\, -8$} &  150 & 2 \\
Taurus	      &  {$165\,$} & {$<\ell<$} & {$\,178$} & {$-20\,$} & {$<b<$} & {$\,-10$} &  153 & 3 \\
Hercules	  &  {$ 40\,$} & {$<\ell<$} & {$\,50$ } & {$  7\,$} & {$<b<$} & {$\, 11$} &  227 & 1 \\
Perseus	      &  {$155\,$} & {$<\ell<$} & {$\,163$} & {$-25\,$} & {$<b<$} & {$\,-15$} &  240 & 3 \\
Orion B	      &  {$201\,$} & {$<\ell<$} & {$\,210$} & {$-17\,$} & {$<b<$} & {$\,-10$} &  423 & 1 \\
Orion A	      &  {$204\,$} & {$<\ell<$} & {$\,218$} & {$-21\,$} & {$<b<$} & {$\,-17$} &  432 & 1 \\
California	  &  {$155\,$} & {$<\ell<$} & {$\,165$} & {$-13\,$} & {$<b<$} & {$\, -5$} &  450 & 3 \\
Mon OB1	      &  {$197\,$} & {$<\ell<$} & {$\,206$} & {$ -1\,$} & {$<b<$} & {$\,  4$} &  745 & 1 \\
Mon R2	      &  {$210\,$} & {$<\ell<$} & {$\,218$} & {$-15\,$} & {$<b<$} & {$\,-10$} &  778 & 1 \\
Cepheus South &  {$105\,$} & {$<\ell<$} & {$\,115$} & {$ 11\,$} & {$<b<$} & {$\, 15$} &  900 & 4 \\
Canis Major	  &  {$220\,$} & {$<\ell<$} & {$\,228$} & {$ -4\,$} & {$<b<$} & {$\,  0$} & 1150 & 5 \\ [0.05ex]
\bottomrule
\end{tabular}
\tiny
\tablefoottext{a}{Distance references: (1) \citet{Zucker2019}; (2) \citet{Boulanger1998}; (3) \citet{Lombardi2010}; 
                  (4) \citet{Schlafly2014}; (5) \citet{Lombardi2011}.}
\end{table}

\section{Method} \label{sect:method}

The relationship between $\SEy$ and $\SEx$ is often complex and cannot be adequately 
described by a single linear model. The observed trend frequently exhibits changes in slope, which are indicative of shifts in the 
underlying physical or chemical processes, such as a transition from a growth phase to one of saturation or destruction in denser 
regions of a cloud. To quantitatively characterize these transitions without presupposing a specific complex physical model, we 
employ a piecewise linear model (PLM). While the true relationship is likely continuous and non-linear, the PLM provides a robust 
method for approximating the average slope within distinct extinction regimes. These fitted slopes can then be used to compare the 
DIB skin effect across different clouds and to infer how environmental conditions may influence the properties of the DIB carrier.

\subsection{Piecewise linear model (PLM)} \label{subsect:PLM}

The fundamental idea of the PLM is to approximate the continuous, curved relationship in log-log space with a series of connected 
straight-line segments. Each linear segment corresponds to a distinct regime where the relationship between $\Wgaia/\Av$ and $\Av$ 
can be approximated by a single power-law. The connection points between segments, referred to as ``knots'', represent the transitional 
extinction values where the nature of this relationship changes. The PLM is a continuous function whose slope changes at these 
discrete knots. For a model with $n$ linear segments defined by $n-1$ knots, the function is expressed as:

\begin{equation}
    f(x) = C + \alpha_1 \cdot x + \sum_{i=1}^{n-1}  \Delta \alpha_i \cdot (x - k_i) \cdot H(x - k_i),
\end{equation}

\noindent where $x = \SEx$ and $f(x) = \SEy$. $C$ is the intercept of the first linear segment. $\alpha_1$ is the slope of the initial 
segment, valid for $x\,{<}\,k_1$. $k_i$ is the position of the $i$-th knot along the x-axis, with the knots ordered such that $k_1 
< k_2 < \dots < k_{n-1}$. Each knot represents a transitional $\SEx$ value. $\Delta \alpha_i$ is the change in the slope at knot $k_i$.
The slope of the $(i\,{+}\,1)$-th segment (between knots $k_i$ and $k_{i+1}$) is the cumulative sum of all preceding slope changes: 
$ \alpha_{i+1} = \alpha_1 + \sum_{j=1}^{i} \Delta \alpha_j$. $H(x-k_i)$ is the Heaviside step function, defined as 0 for $x\,{<}\,k_i$ 
and 1 for $x\,{\geqslant}\,k_i$.

The PLM serves as a powerful empirical tool for analyzing trends in the data. However, it is important to recognize its inherent 
assumptions. The model approximates a potentially smooth, continuous curve with discrete transitions at the knots. In physical 
reality, these transitions are likely more gradual. Therefore, the knots should be interpreted not as precise physical thresholds, 
but as the characteristic extinction values around which the dominant physical regime changes. Similarly, the linearity of each 
segment is an approximation of the true relationship within that regime.

\subsection{Fitting Procedure} \label{subsect:fit}

A key challenge in applying a PLM is selecting the appropriate number of linear components ($n_p$). A model with too few components 
may fail to capture significant trends in the data, while a model with too many free parameters risks overfitting, where it begins 
to model random statistical noise rather than the underlying physical relationship. To balance model complexity and goodness-of-fit, 
we adopted a systematic approach. For each target cloud, we fit a series of PLMs with an increasing number of components ($n_p\,{=}\,1{-}4$)
and then selected the optimal model by visual inspection. This ensures that our final results reflect genuine, large-scale changes 
in the DIB behavior.

For our fitting procedure, the model with $n_p$ components is parameterized directly by $\{C,\ \alpha_1,\ \Delta \alpha_i,\ k_i\}$,
where $i$ ranges from $1$ to $n_p-1$. This parameterization, which models the changes in slope ($\Delta \alpha_i$) at each knot, is 
often more statistically robust and leads to more efficient sampling of the parameter space compared to fitting the absolute slopes 
directly. We assigned uniform priors to the model parameters. The specific ranges were: $C \in [0, 5]$, $\alpha_1 \in 
[-5, 5]$, and $\Delta \alpha_i \in \left[-15, 15\right]$. The priors for the knot locations, $k_i$, were also uniform, constrained 
to lie within the observed range of $\SEx$ for each cloud and ordered such that $k_1 < k_2 < \dots < k_{n_p-1}$.

To properly account for observational uncertainties in both $\SEx$ and $\SEy$, we constructed a likelihood function following the 
methodology of \citet{Li2020}. The total likelihood for a set of $N$ data points $D\,{=}\,\{x_i, y_i\}$ given a model prediction 
$M\,{=}\,\{x_{m},y_{m}\}$ is the product of the individual probabilities:

\begin{align}
    \mathcal{L}(D|M) &= \prod_{i=1}^{N} P(D_i|M), \\
    P(D_i|M) &= \frac{1}{2\pi\sigma_{x_i}\sigma_{y_i}} 
    \exp\left[-\frac{1}{2}\left(
        \frac{(x_i - x_m)^2}{\sigma_{x_i}^2} 
        + \frac{(y_i - y_m)^2}{\sigma_{y_i}^2}
    \right)\right],
\end{align}

\noindent where the uncertainties $\sigma_x$ and $\sigma_y$ on the logarithmic quantities are derived from the observational 
errors via standard error propagation:

\begin{align}
    \sigma_x &= \frac{\sigma(\Av)}{\ln(10) \cdot \Av}, \\
    \sigma_y &= \frac{1}{\ln(10)}\left( \left[\frac{\sigma(\Wgaia)}{\Wgaia}\right]^2 + \left[\frac{\sigma(\Av)}{\Av}\right]^2\right)^{1/2}.
\end{align}

The posterior probability distributions for the model parameters were sampled using the nested sampling algorithm \citep{Skilling2004}, 
as implemented in the Python package \texttt{dynesty} \citep{Speagle2020}. From the resulting posterior distributions, the best-fit 
value for each parameter is taken as the median (50th percentile), and the uncertainties are defined by the 16th and 84th percentiles, 
which enclose a 68\% ($1\sigma$) confidence interval.

While the model is fit using the slope changes, $\Delta \alpha_i$, the absolute slopes of each segment, $\alpha_i$, are more physically 
intuitive. Therefore, for our final analysis and reporting in Tables \ref{tab:plm} and \ref{tab:plm-full}, we calculate the absolute 
slopes from the posterior samples using the relation $\alpha_{i+1} = \alpha_1 + \sum_{j=1}^{i} \Delta \alpha_j$. This allows for a 
direct assessment of the DIB behavior in each regime, such as identifying the onset of saturation where the slope approaches a value 
of $\alpha\,{=}\,{-}1$.

\begin{figure*}
  \centering
  \includegraphics[width=16cm]{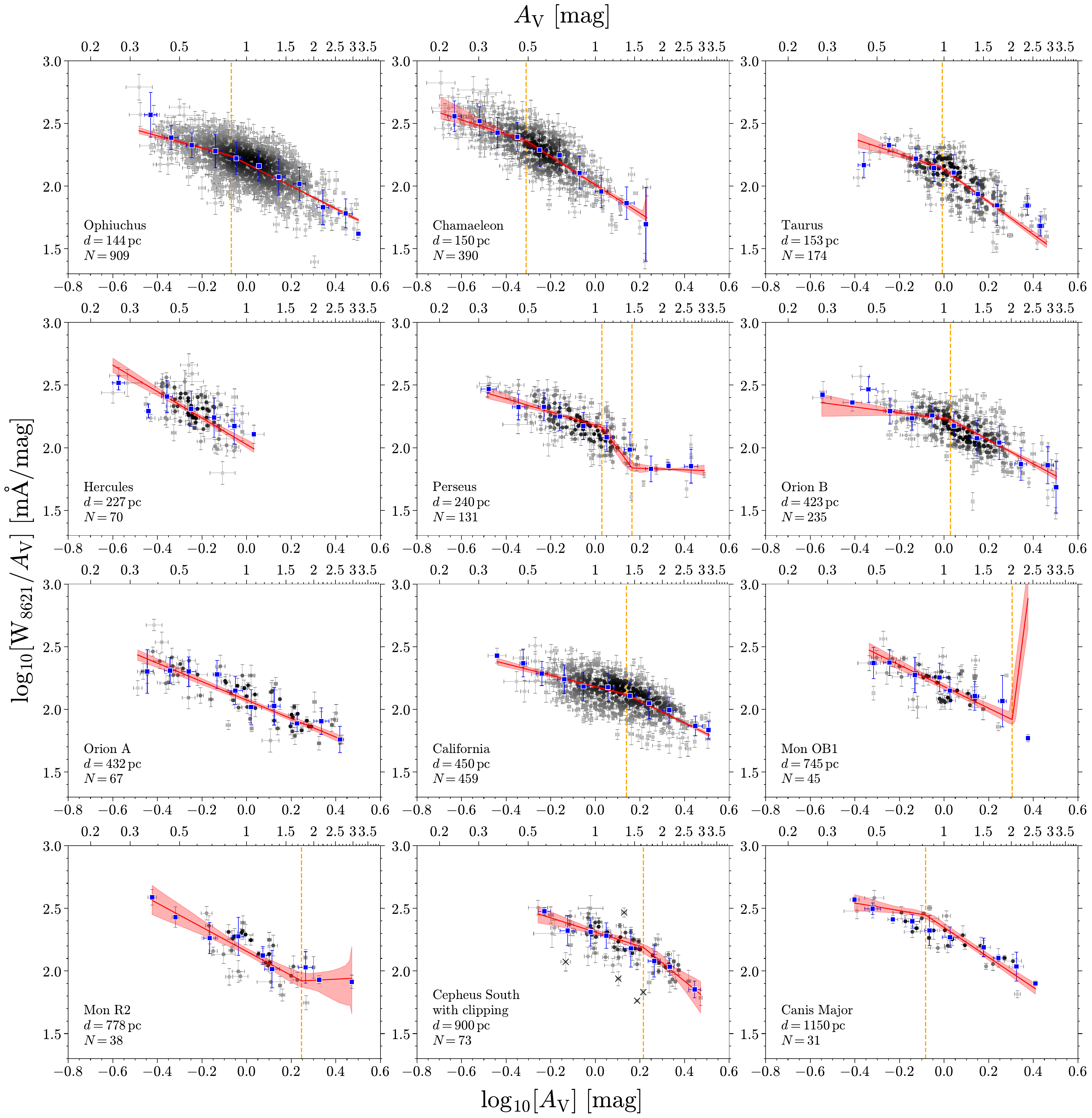}
  \caption{Variation of $\SEy$ as a function of $\SEx$ for each target cloud. The corresponding $\Av$ scale is shown on the top 
  of each panel. Points are color-coded by their number density, estimated via a Gaussian KDE. Blue squares represent the median 
  $\SEy$ in bins of $\SEx$ (from --0.8 to 0.6 with a step of 0.1). The solid red line and shaded region show the selected Piecewise 
  Linear Model (PLM) and its 99.7\% credible interval, respectively. Vertical dashed orange lines indicate the knot locations (see 
  Sect. \ref{subsect:PLM}). 
  The full PLM fitting results with $n_p\,{=}\,1{-}4$ for each cloud can be found in Fig. \ref{fig:mc-plm-all}.
  For Cepheus South, points marked with black crosses were clipped as outliers and excluded from the fit
  (see Sect. \ref{sect:results} for details).
  Each panel is labeled with the cloud name, distance, and the number of sightlines used.}
  \label{fig:skin-effect}
\end{figure*}

\section{Results} \label{sect:results}

The full PLM fitting results for all target clouds, testing models with $n_p\,{=}\,1$ to $4$ components, are presented in Fig. 
\ref{fig:mc-plm-all}. The complete set of derived parameters for every tested model is listed in Table \ref{tab:plm-full}. In the 
figures, the solid red line indicates the posterior median model, while the shaded red region represents the 99.7\% 
(3$\sigma$-equivalent) credible interval, derived from the full posterior distribution of the model parameters.

For each cloud, we selected the optimal model from the set ($n_p\,{=}\,1{-}4$) based on visual inspection, considering the physical 
plausibility of the resulting parameters. First, we avoided PLMs with unrealistically large $\alpha$ values (e.g., $\alpha_2$ 
in the $n_p\,{=}\,3$ model for 
Ophiuchus), which imply an abrupt change in the variation of $\SEy$ that cannot be reliably represented by a PLM and would distort 
the fitting of other slopes. Second, for clouds with fewer data points (e.g., Hercules), the fitting results are more sensitive to 
scatter. In such cases, we adopted simpler and more stable PLMs with smaller $n_p$ to minimize overfitting. The selected PLM for 
each cloud is shown in Fig. \ref{fig:skin-effect}, with the corresponding parameters listed in Table \ref{tab:plm}.
A brief summary of the model selection for each cloud is provided below.

{\it Ophiuchus, Chamaeleon, Orion B,} and {\it California:} These four clouds exhibited similar behavior. While models with $n_p\,{>}\,2$ 
were tested, the more complex models tended to introduce abrupt, high-magnitude slope changes to accommodate minor scatter, which 
were not statistically justified. We therefore conservatively selected the $n_p\,{=}\,2$ model for all four, characterizing the 
trend as a simple break.

{\it Taurus} is the only cloud in our sample where the median trend suggests an initial ascent in $\SEy$ at low $\Av$, consistent 
with the predictions of \citetalias{Sonnentrucker1997}. However, due to the small number of data points in this regime 
($\Av\,{\lesssim}\,0.6$\,mag), the PLM fit was unable to robustly capture this feature. A test fit required $n_p\,{\geqslant}\,8$ 
to model the ascent, which is an unjustifiably complex model that constitutes severe overfitting. We selected $n_p\,{=}\,2$ to 
describe the dominant trend of decreasing $\SEy$ after its peak.

{\it Hercules} has the lowest line-of-sight dust column in our sample ($\Av\,{\sim}\,1$\,mag) and exhibits significant scatter. 
Although a few points at $\Av\,{<}\,0.3$ mag have a lower $\SEy$, the data are too sparse to confirm a distinct trend. We selected 
the single-component model ($n_p\,{=}\,1$) as adding more components were merely fitting noise.

{\it Perseus:} 
For $\Av\,{>}\,1.5$\,mag, a distinct flattening end is evident in both the median trend and the models with 
$n_p\,{\geqslant}\,2$. This suggests a potential re-increase of $\Wdib$ with $\Av$ in dense regions--a newly observed phenomenon. 
We finally selected the $n_p\,{=}\,3$ model because it best captures the behavior across both low- and high-extinction regimes, 
matching to the median trend, although the data in these regime is limited to a small number of sightlines exhibiting large 
scatter.
The fit reveals three distinct regimes:
An initial decline with slope $\alpha_1={-}0.53_{{-}0.04}^{+0.04}$ for $\Av\,{\lesssim}\,1$\,mag. A dramatically steeper drop 
with $\alpha_2\,{=}\,{-}2.40_{{-}0.25}^{+0.23}$ between $\Av\,{\approx}\,1{-}1.5$ mag. And a flattening to a near-zero slope 
($\alpha_3\,{=}\,{-}0.06_{{-}0.46}^{+0.45}$) for $\Av\,{>}\,1.5$\,mag.

{\it Orion A:} A simple linear model ($n_p\,{=}\,1$) was sufficient to describe the trend in Orion A, yielding a continuous decrease 
with slope $\alpha_1\,{=}\,{-}0.74_{{-}0.02}^{+0.02}$. While the median trend appears flatter for $\Av\,{<}\,0.8$ mag, the PLM fit, 
which accounts for all individual data points and their uncertainties, finds no statistical support for a break. The $n_p\,{=}\,3$ 
model for Orion A (see Fig. \ref{fig:mc-plm-all}) shows a significant deviation between the median model line and the credible region, 
a clear sign of a poorly constrained and non-robust fit.

{\it Mon OB1:} For this cloud, the simple $n_p\,{=}\,1$ model failed to capture the global trend, while models with $n_p\,{>}\,1$ 
primarily used the extra components to fit scattered points at $\Av\,{>}\,1.5$ mag. The $n_p\,{=}\,2$ model was statistically preferred, 
but the second slope, $\alpha_2$, was poorly constrained. Therefore, we selected the $n_p\,{=}\,2$ model but only consider the first 
slope, $\alpha_1\,{=}\,{-}0.86_{{-}0.05}^{+0.05}$, to be a physically robust result.

{\it Mon R2:} Similar to Perseus, Mon R2 exhibits a plateau after an initial decline. The trend is best described by the $n_p\,{=}\,2$ 
model, with a steep initial slope of $\alpha_1\,{=}\,{-}0.96_{{-}0.09}^{+0.09}$ followed by a flat second segment where 
$\alpha_2\,{=}\,0.09_{{-}0.47}^{+0.96}$.

{\it Cepheus South:} Initial fits to this cloud failed for all tested $n_p$ values. 
The failure was driven by two outlier data points 
at $\SEx\,{\sim}\,0.2$ and $\SEy\,{\sim}\,1.8$, which had very small reported uncertainties and thus exhibited a high impact on the 
fit. As this was the only cloud where all tested fits failed, we took an exceptional step of applying an outlier clipping procedure 
based on a Gaussian kernel density estimation (five clipped points are marked by black crosses in Figs.
\ref{fig:skin-effect} and \ref{fig:mc-plm-all}). After clipping, the data were well-described by an $n_p\,{=}\,2$ model. 
We refrained from applying this clipping 
to other clouds, as scattered points may contain real physical information.

{\it Canis Major:} The $n_p\,{=}\,2$ model was selected. It describes a trend that begins with a much flatter slope 
($\alpha_1\,{=}\,{-}0.30_{{-}0.07}^{+0.07}$) and then steepens to $\alpha_2\,{=}\,{-}1.20_{{-}0.16}^{+0.16}$.

\begin{figure}
  \centering
  \includegraphics[width=8cm]{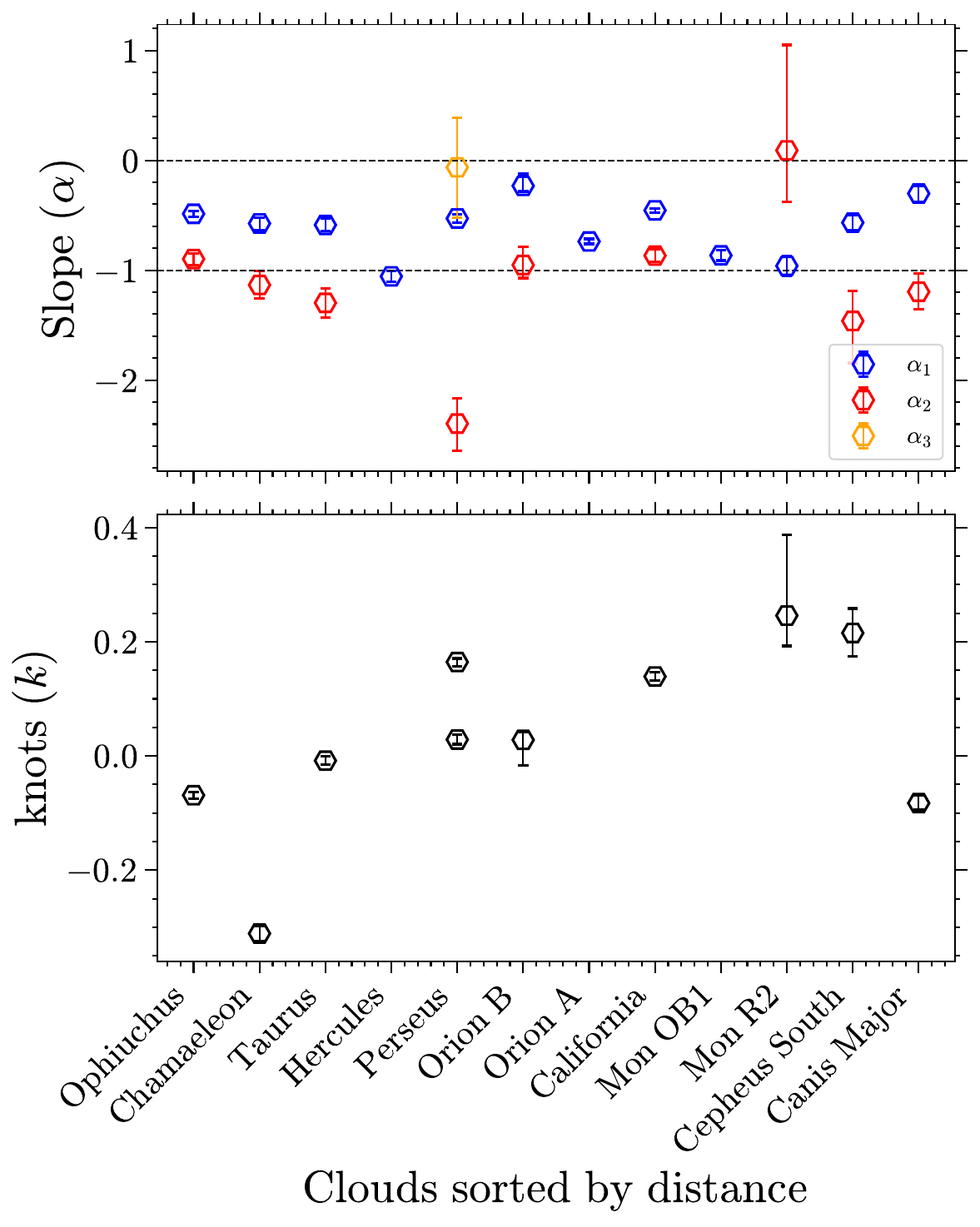}
  \caption{Comparison of the slopes and knots of the selected PLM between target clouds. The slopes are arranged with increasing
  $\SEx$ and marked with different colors.}
  \label{fig:compare}
\end{figure}

Our model selection indicates a clear preference for simpler models, constrained by both observational scatter and inherent model 
limitations. Eight clouds are best described by a two-component model ($n_p\,{=}\,2$), three\footnote{Including Mon OB1, for which 
only $\alpha_1$ is valid in the $n_p\,{=}\,2$ PLM fit.} by a single-component model ($n_p\,{=}\,1$), and only one, Perseus, requires 
a three-component model ($n_p\,{=}\,3$). A comparison of the fitted slopes and knot locations for the selected models is presented 
in Fig. \ref{fig:compare}. Most initial slopes ($\alpha_1$; blue hexagons) fall between $0$ and ${-}1$, indicating a general decline 
of $\SEy$ with respect to $\SEx$ as sightlines probe from diffuse to translucent regions. Clouds fitted with an $n_p\,{=}\,2$ model 
typically exhibit a steepening trend, where the slope transitions from a shallower value (${-}1\,{<}\,\alpha_1\,{<}\,0$) in the 
outer regions to a steeper one ($\alpha_2\,{\lesssim}\,{-}1$) in the inner regions. This is qualitatively consistent with the 
predictions of the \citetalias{Sonnentrucker1997} PIE model. Two notable exceptions deviate from this simple steepening pattern: 
both Perseus and Mon R2 display a distinct flattening of the slope ($\alpha\,{\approx}\,0$) in their densest regions 
($\Av\,{\gtrsim}\,1.5$\,mag), a feature not explicitly predicted by the PIE model.

\begin{table*}
\begin{threeparttable}
\renewcommand{\arraystretch}{1.5}
\caption{Results of the selected PLM for each cloud. }
\small
\centering
\label{tab:plm}
\begin{tabular}{lrcccc} 
\toprule
Name &  \multicolumn{1}{c}{$N_{\rm los}$} & $n_p$ & $C$ & $\alpha$ & $k$ \\ 
\midrule
Ophiuchus  & 909 & 2 & $2.21_{-0.01}^{+0.01}$ &  $\alpha_1=-0.49_{-0.02}^{+0.02}$, $\alpha_2=-0.90_{-0.05}^{+0.05}$ & $k_1=-0.07_{-0.01}^{+0.01}$ \\ 
Chamaeleon & 390 & 2 & $2.18_{-0.02}^{+0.02}$ & $\alpha_1=-0.57_{-0.06}^{+0.05}$, $\alpha_2=-1.13_{-0.12}^{+0.12}$ & $k_1=-0.31_{-0.01}^{+0.01}$ \\ 
Taurus	   & 174 & 2 & $2.14_{-0.01}^{+0.01}$ & $\alpha_1=-0.59_{-0.06}^{+0.06}$, $\alpha_2=-1.30_{-0.13}^{+0.13}$ & $k_1=-0.01_{-0.01}^{+0.01}$ \\ 
Hercules   &  70 & 1 & $2.03_{-0.01}^{+0.01}$ & $\alpha_1=-1.06_{-0.05}^{+0.05}$ & ... \\ 
Perseus	   & 131 & 3 & $2.18_{-0.01}^{+0.01}$ & $\alpha_1=-0.53_{-0.04}^{+0.04}$, $\alpha_2=-2.40_{-0.25}^{+0.23}$, $\alpha_3=-0.06_{-0.46}^{+0.45}$ & $k_1=0.03_{-0.01}^{+0.01}$, $k_2=0.16_{-0.01}^{+0.01}$ \\ 
Orion B	   & 235 & 2 & $2.23_{-0.01}^{+0.02}$ & $\alpha_1=-0.23_{-0.05}^{+0.11}$, $\alpha_2=-0.95_{-0.12}^{+0.17}$ & $k_1=0.03_{-0.04}^{+0.01}$ \\ 
Orion A	   &  67 & 1 & $2.07_{-0.01}^{+0.01}$ & $\alpha_1=-0.74_{-0.02}^{+0.02}$ & ... \\ 
California & 459 & 2 & $2.18_{-0.01}^{+0.01}$ & $\alpha_1=-0.46_{-0.02}^{+0.02}$, $\alpha_2=-0.87_{-0.05}^{+0.05}$ & $k_1=0.14_{-0.01}^{+0.01}$ \\ 
Mon OB1	   &  45 & 2 & $2.18_{-0.01}^{+0.01}$ & $\alpha_1=-0.86_{-0.05}^{+0.05}$, $\alpha_2=13.65_{-0.70}^{+0.40}$ & $k_1=0.31_{-0.01}^{+0.01}$ \\ 
Mon R2	   &  38 & 2 & $2.16_{-0.01}^{+0.01}$ & $\alpha_1=-0.96_{-0.09}^{+0.09}$, $\alpha_2=0.09_{-0.47}^{+0.96}$  & $k_1=0.25_{-0.05}^{+0.14}$ \\ 
Cepheus South & 73 & 2 & $2.31_{-0.01}^{+0.01}$ & $\alpha_1=-0.57_{-0.07}^{+0.07}$, $\alpha_2=-1.46_{-0.38}^{+0.27}$ &
$k_1=0.22_{-0.04}^{+0.04}$ \\
Canis Major	  & 31 & 2 & $2.42_{-0.01}^{+0.01}$ & $\alpha_1=-0.30_{-0.07}^{+0.07}$, $\alpha_2=-1.20_{-0.16}^{+0.16}$ & $k_1=-0.08_{-0.01}^{+0.01}$ \\ 
\bottomrule
\end{tabular}
\begin{tablenotes}
    \tiny
    \item[] Columns: 
    \item[1] $N_{\rm los}$: Number of sightlines with $\Wgaia$ and $\Av$ measurements in each cloud.
    \item[2] $n_p$: Number of linear segments for the selected PLM.
    \item[3] $C$: intercept.
    \item[4] $\alpha$: fitted and calculated slopes of each linear segment in PLM ($ \alpha_{i+1} = \alpha_1 + \sum_{j=1}^{i} \Delta \alpha_j $
when $i>1$)
    \item[5] $k$: fitted knot position in PLM. 
    \item[] Notes: 
    \item[i] The PLM fit to Cepheus South is with outlier clipping (see Sect. \ref{sect:results}). 
    \item[ii] For Mon OB1, only $\alpha_1$ is treated as a reasonable and valid fitted slope.
\end{tablenotes}
\end{threeparttable}
\end{table*}

\section{Discussion} \label{sect:discuss}

\subsection{Taurus region and carrier ionization properties} \label{subsect:taurus}

Among the clouds in our sample, Taurus is the only one that exhibits a clear ascending trend of $\SEy$
at low extinction ($\Av\,{\lesssim}\,0.6$\,mag), a behavior predicted by the \citetalias{Sonnentrucker1997} PIE 
model for an ionized carrier. This makes it an excellent target for studying the properties of 
the DIB\,$\lambda$8621 carrier within the framework of the PIE model. 
This trend is not seen in other target clouds, though they also contain sightlines with low extinction.
As the increase of $\SEy$ occurs in the UV-penetrated regions, a possible explanation is that sightlines with strong
UV radiation are not probed for these clouds. The ascending trend was seen in Orion region for DIB\,$\lambda$6284
by \citet{Jenniskens1994} but their background stars are not contained in our sample. 

\begin{figure}
  \centering
  \includegraphics[width=8cm]{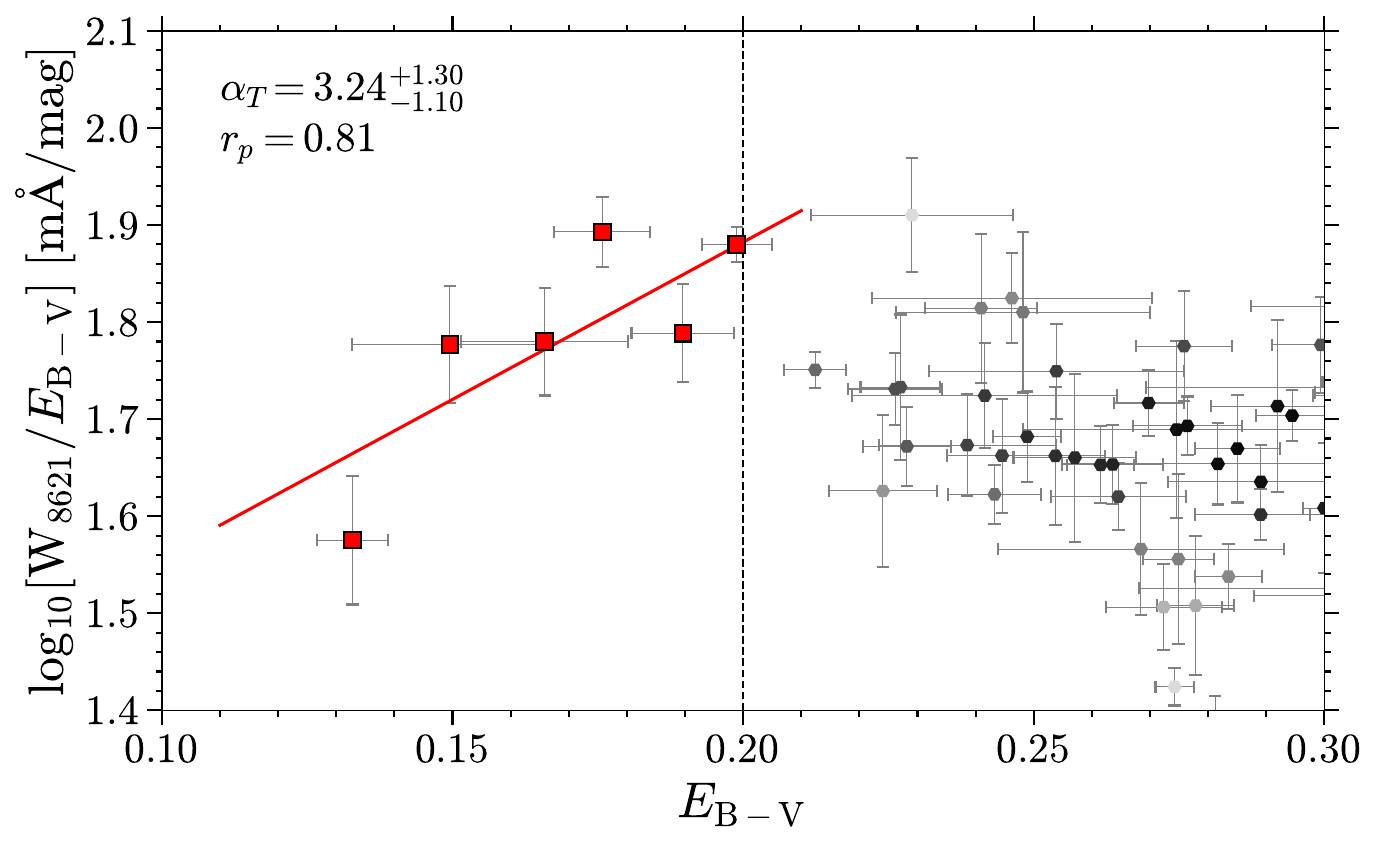}
  \caption{Linear fit to the relationship between ${\rm log_{10}}(\Wgaia/\EBV)$ and $\EBV$ for sightlines in the Taurus cloud with 
  $\EBV\,{<}\,0.2$\,mag. The data points included in the fit are shown as red squares. The best-fit line is shown in red, and the 
  resulting slope ($\alpha_T$) and the Pearson correlation coefficient ($r_p$) are indicated.}
  \label{fig:Taurus}
\end{figure}

\begin{figure}
  \centering
  \includegraphics[width=8cm]{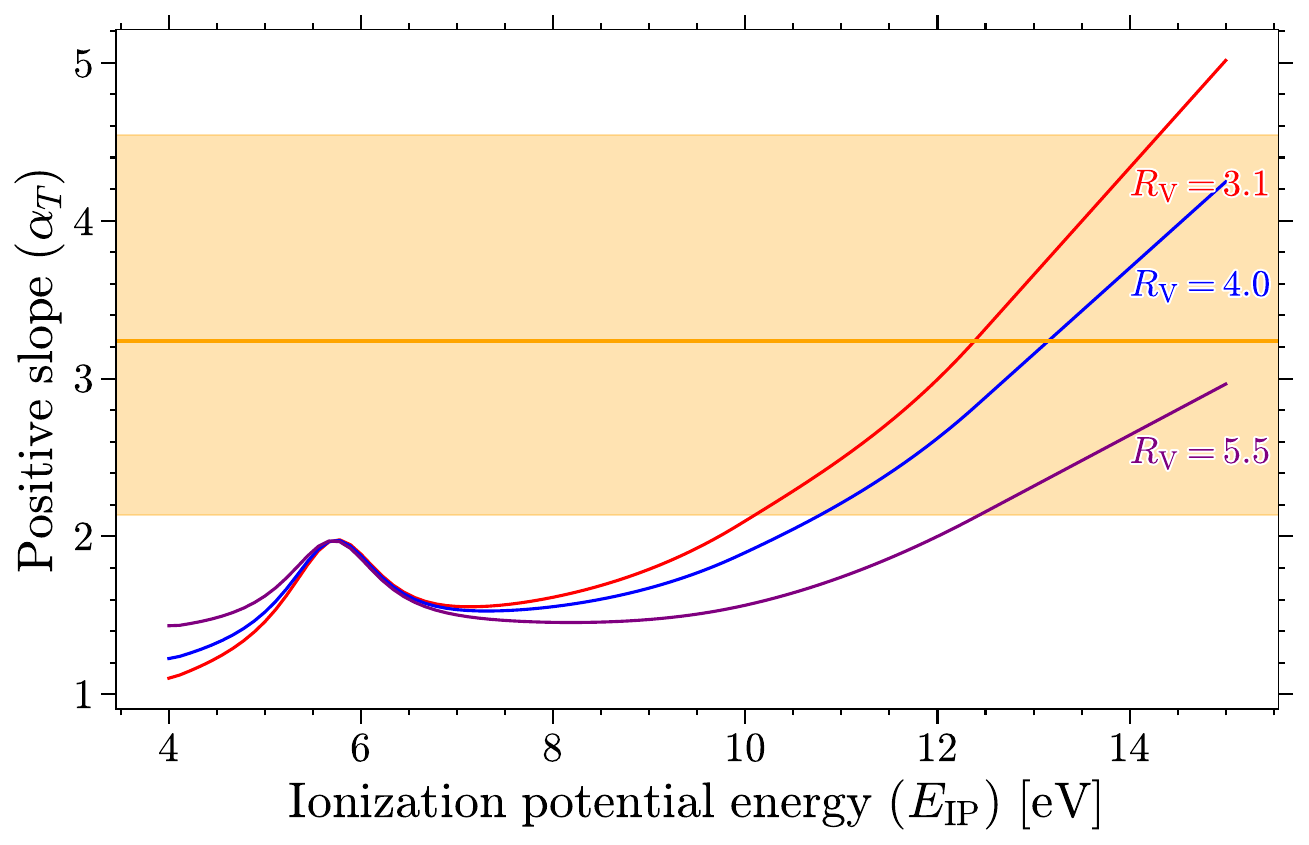}
  \caption{The theoretical relationship between the ionization potential ($E_{\rm IP}$) of a DIB carrier and the expected 
  observational slope ($\alpha_T$) from the PIE model of \citetalias{Sonnentrucker1997}. The relationship is plotted for three 
  different $\Rv = 3.1, 4.0,$ and $5.5$. The vertical orange line and shaded region represent our measured value of $\alpha_T$ 
  for DIB\,$\lambda$8621 (from Fig. \ref{fig:Taurus}) and its 1$\sigma$ uncertainty, respectively. The intersection of our 
  measurement with the $\Rv\,{=}\,3.1$ curve implies $E_{\rm IP}\,{\approx}\,12$\,eV.}
  \label{fig:PIE}
\end{figure}

By simplifying the relation between $\Wdib$ and optical depth and considering a homogeneous diffuse region,
the positive slope in Taurus is expressed as $\alpha_T\,{=}\,\frac{A_\lambda}{5\EBV}\,{=}\,\frac{A_\lambda 
\cdot \Rv}{5\Av}$ (Eq. 8 in \citetalias{Sonnentrucker1997}).
We performed a linear fit to the sightlines in Taurus with $\EBV\,{<}\,0.2$\,mag and obtained a slope of 
$\alpha_T\,{=}\,3.24_{{-}1.10}^{+1.30}$, with uncertainties estimated via Monte Carlo resampling (shown in Fig. \ref{fig:Taurus}). 
Within the PIE model, the value of $\alpha_T$ depends on the characteristic wavelength of the ambient UV radiation field responsible 
for ionizing the DIB carriers. For a given $\Rv$, this allows one to determine the corresponding photon energy ($h\nu$), which is 
interpreted as the $E_{\rm IP}$ of the carrier. Figure \ref{fig:PIE} illustrates the theoretical dependence of $\alpha_T$ on 
$E_{\rm IP}$ for several values of $\Rv$. Our measured $\alpha_T$ corresponds to an ionization potential of 
$E_{\rm IP}\,{=}\,12.40^{+1.90}_{-2.29}$\,eV for the $\lambda$8621 carrier when applying $\Rv\,{=}\,3.1$.
The $E_{\rm IP}$ uncertainty is simply estimated by the fitting error of $\alpha_T$. 
Our calculations of this relationship 
are consistent with those presented in \citetalias{Sonnentrucker1997} (their Table 2); minor differences arise from our use of the 
\citet{CCM89} extinction curve versus the \citet{Savage1977} curve used in their study. An $E_{\rm IP}$ of $\sim$12\,eV is compatible 
with the known second ionization potentials of PAHs \citep{Malloci2007,Holm2011} and fullerenes (e.g., $C_{20}$--$C_{70}$; 
\citealt{Diaz-Tendero2006}). This suggests that the $\lambda$8621 carrier could plausibly be a singly-charged PAH or fullerene 
cation, an inference analogous to that drawn by \citetalias{Sonnentrucker1997} for the optical DIBs $\lambda$5780, $\lambda$5797, 
and $\lambda$6614.

The skin effect shows that DIB carriers are most abundant in the outer, UV-illuminated layer of clouds. Consequently, the relative 
distribution of different DIBs as a function of cloud opacity--a concept known as the spatial sequence--is a key diagnostic for 
investigating carrier properties. \citetalias{Sonnentrucker1997} argued that a DIB whose normalized strength ($\Wdib/\EBV$) peaks 
at a lower value of $\EBV$ must have a carrier that is more resistant to the stronger UV radiation field present in diffuse 
environments. This framework establishes a sequence of DIBs from those favoring diffuse regions to those favoring denser, more 
shielded regions: $\lambda5780 \rightarrow \lambda6614 \rightarrow \lambda5797 \rightarrow \lambda6379$. This ordering is consistent 
with sequences derived from other diagnostics, such as the tightness of the correlation between $\Wdib$ and {\ion{H}{i}} column 
density \citep{Friedman2011} and the peak of $\Wdib/\EBV$ relative to the molecular hydrogen fraction \citep{Fan2017}, although 
these latter studies did not include $\lambda$6379.

Our results for DIB\,$\lambda$8621, which peaks at $\EBV\,{\sim}\,0.2$\,mag in Taurus (see Fig. \ref{fig:Taurus}), place it between 
$\lambda$5780 and $\lambda$6614 
in this spatial sequence. This placement suggests that $\lambda$8621 behaves as a $\sigma$-type DIB, which are known to correlate 
well with $\lambda$5780, a finding consistent with the suggestion by \citet{Wallerstein2007}. Interestingly, this spatial sequence 
is also consistent with the descending order of their average $\Wdib/\EBV$ measured along sightlines toward field stars \citep{Fan2019}: 
$W_{5780}/\EBV > W_{8621}/\EBV > W_{6614}/\EBV > W_{5797}/\EBV > W_{6379}/\EBV$. This consistency supports the physical picture where 
carriers more stable against UV photons (e.g., the $\lambda$5780 carrier) exhibit higher relative strengths in low-density, high-UV 
environments compared to less stable carriers (e.g., the $\lambda$5797 carrier) that are more rapidly destroyed in such conditions 
\citep{KW1988,Herbig1995,Cami1997,Vos2011}. Furthermore, this sequence may also relate to carrier size. An analysis of DIB profile 
substructures by \citet{MacIsaac2022} estimated an increasing carrier size (in number of carbon atoms) along the sequence 
$\lambda6614 < \lambda5797 < \lambda6379$, assuming linear or spherical molecules. Taken together, these converging lines of evidence
underscore that the behavior of DIBs as a function of dust extinction is a powerful diagnostic tool for exploring the physical 
and chemical nature of their carriers.

\subsection{Implications of environmental conditions in different clouds} \label{subsect:cloud}

In nearly all the clouds investigated, the normalized DIB strength, $\SEy$, decreases with increasing extinction, $\SEx$, exhibiting 
an average slope ($\alpha_1$) between 0 and --1. Besides Taurus, Hercules is another exception, showing a flatter stage at $\Av\,{\sim}\,0.3$\,mag 
before steepening. Such a negative slope in this log-log plot signifies that while $\Wgaia$ increases as sightlines probe deeper into a 
cloud, it does so at a slower rate than $\Av$. According to the \citetalias{Sonnentrucker1997} PIE model, the abundance of an 
ionized carrier is proportional to the local $\UVflux$ and should therefore decrease exponentially as the UV field is attenuated 
deeper inside a cloud. To explain the observed net increase in $\Wgaia$ (despite slower than $\Av$), at least two effects 
must be considered. First, the accumulation of gas toward the cloud center increases the reservoir of neutral precursor molecules 
available to be ionized. Second, the cloud's geometry plays a role; for instance, in a simple spherical model, the line-of-sight 
path length through the carrier-bearing region increases toward the cloud center, boosting the total column density to which 
$\Wgaia$ is proportional.

The variation of the slope $\alpha_1$ among different clouds likely reflects differences in their internal structures, such as their 
gas and dust density profiles. A more rapid accumulation of dust, for example, would lead to faster UV attenuation and thus a more 
negative (steeper) slope $\alpha_1$. This decline continues until a critical extinction, $\Av^{C}$, is reached, at which point the 
slope approaches --1. In the log-log representation of this work, a slope of --1 implies that $\Wgaia$ has become constant, 
independent of any further increase in $\Av$. The physical interpretation within the PIE model is that at $\Av\,{>}\,\Av^{C}$, 
the internal UV field is so strongly attenuated that the DIB carrier is no longer produced. 
Our observations reveal that $\Av^{C}$ varies significantly among clouds.
For the Hercules and Mon R2 clouds, their $\alpha_1$ approaches a value of --1, which implies a low $\Av^{C}$ of 0.4--0.5\,mag. 
In contrast, the slopes for Ophiuchus, Orion A, California, and Mon OB1 remain greater than --1 (i.e., shallower) even at 
$\Av\,\sim\,2{-}3$\,mag. Other clouds exhibit slopes that decrease from a value between 0 and --1 to below --1, with this 
transition occurring at a $\Av^{C}$ of approximately 1--2\,mag, a range characteristic of the translucent-to-dense cloud regime (\citealt{SM2006}).

\begin{figure}
  \centering
  \includegraphics[width=8cm]{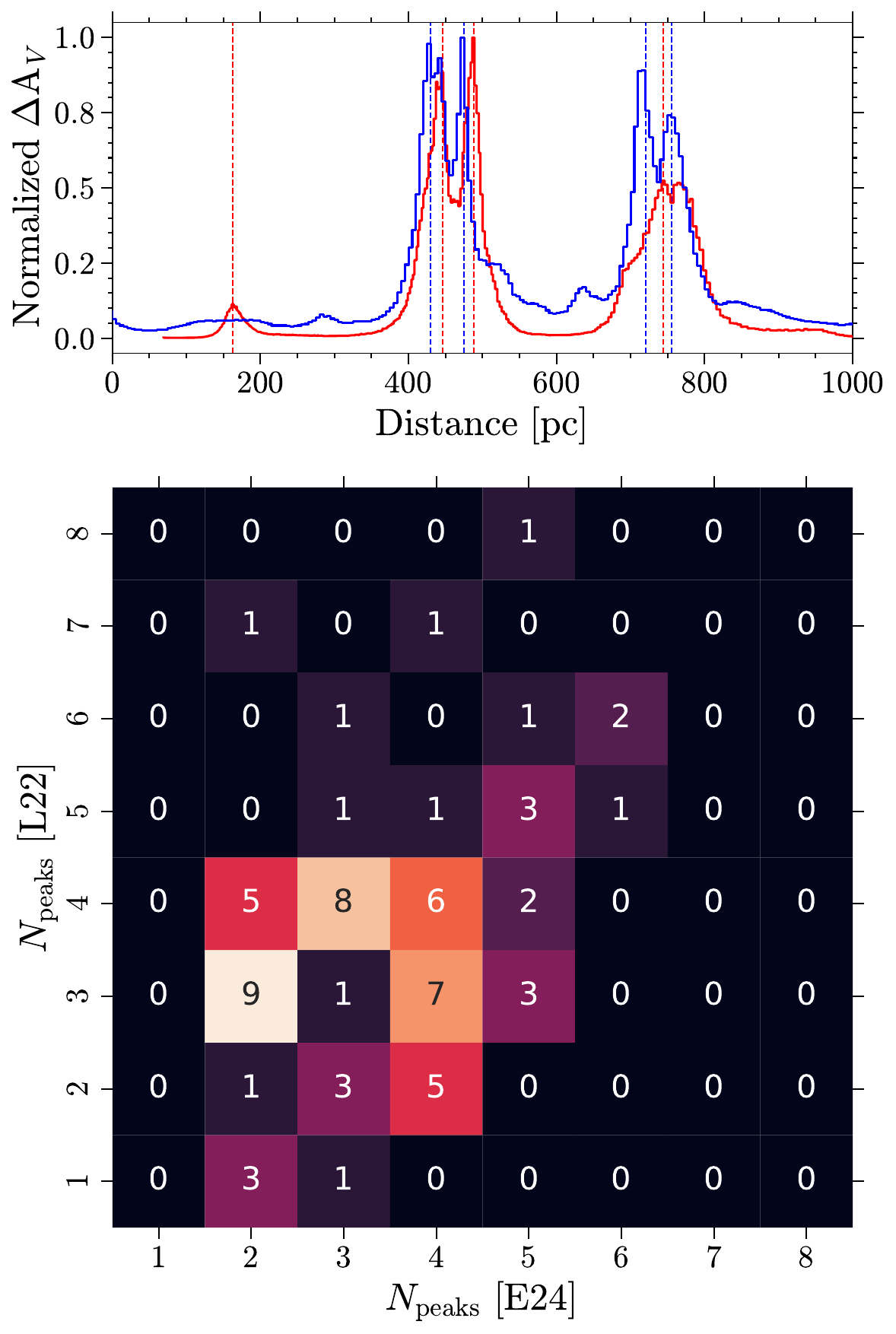}
  \caption{{\it Upper panel:} Normalized dust density profiles toward the sightline of 
  $(\ell,b)=(216.9^{\circ},{-}18.6^{\circ})$ in Orion A (432\,pc),
  with a background star at 936\,pc. The profiles are derived from the dust maps of 
  \citet[][\citetalias{Lallement2022}, blue]{Lallement2022} and
  \citet[][\citetalias{Edenhofer2024}, red]{Edenhofer2024}.
  The dashed vertical lines indicate the detected dust components. 
  {\it Lower panel:} The heatmap of the number of the detected dust components ($N_{\rm peaks}$) 
  for the 67 sightlines in Orion A derived with the dust maps of \citetalias{Lallement2022} 
  and \citetalias{Edenhofer2024}. 
  The color scale and numbers represent that both maps detect a specific number of $N_{\rm peaks}$ 
  (e.g., 6 sightlines have $N_{\rm peaks}\,{=}\,4$ in both \citetalias{Lallement2022} and \citetalias{Edenhofer2024}).}
  \label{fig:peak1}
\end{figure}

\begin{figure*}
  \centering
  \includegraphics[width=16cm]{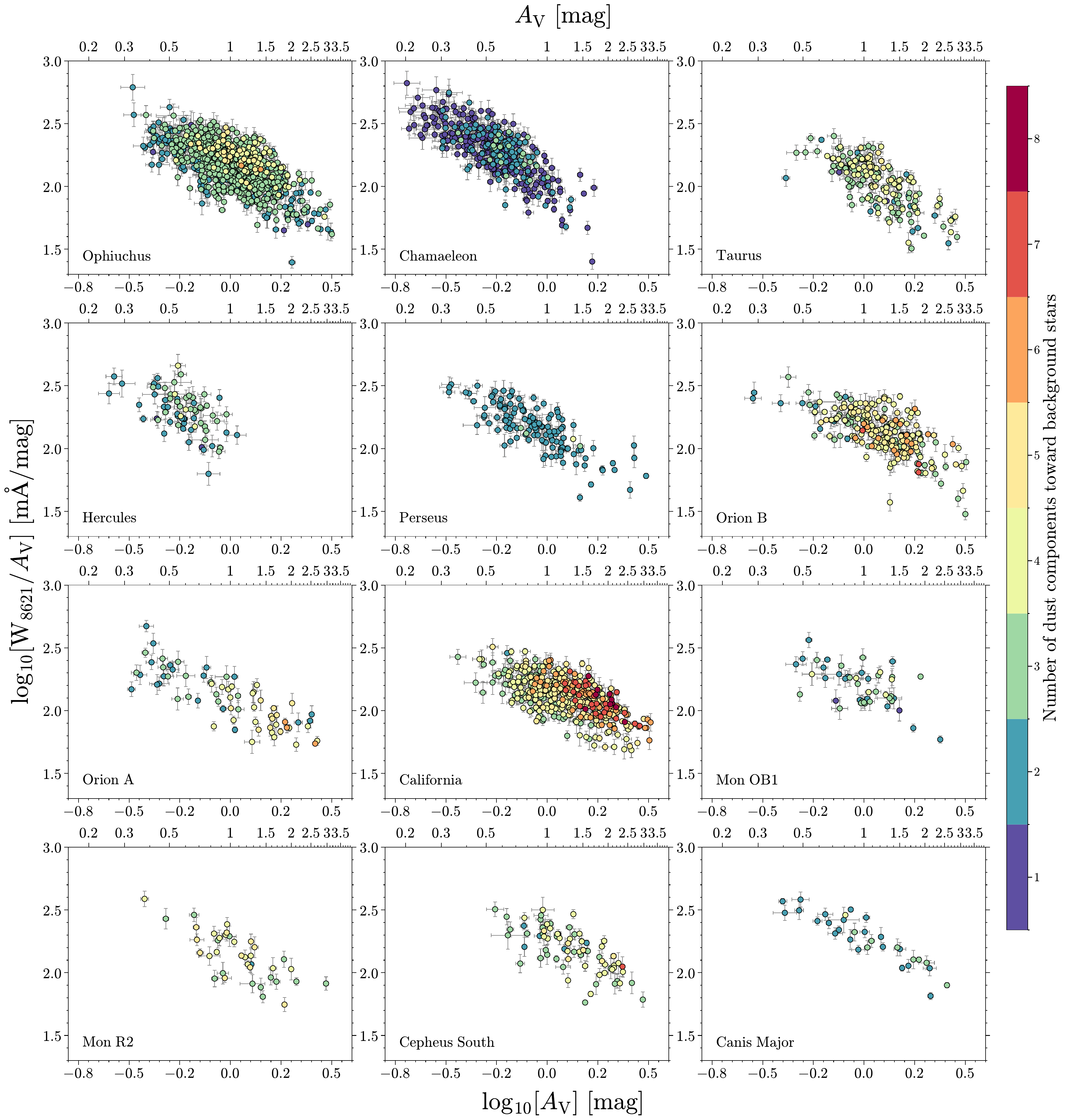}
  \caption{Variation of $\SEy$ as a function of $\SEx$ for each target cloud, color-coded by $N_{\rm peaks}$
  of each sightline derived from the \citetalias{Edenhofer2024} map. }
  \label{fig:peak2}
\end{figure*}

While the strength of the UV radiation field certainly influences the depth to which carriers can survive, it is likely 
insufficient on its own to explain this wide range of $\Av^{C}$. The local electron density ($N_e$), which is nearly constant in 
diffuse region and starts to sharply drop in the transition from diffuse to translucent gas \citep{Neufeld2005,SM2006}, is another 
key factor. A lower $N_e$ reduces the carrier-electron recombination rate, which would counteract the effect of UV attenuation and 
lead to a shallower slope (larger $\alpha$), thereby increasing $\Av^{C}$. 
This hypothesis can be tested when the estimation of $N_e$ is available for enough sightlines.

Furthermore, the clumpy, non-uniform nature of MCs, as revealed by 3D dust maps \citep[e.g.,][]
{Lallement2019,RezaeiKh2020,Dharmawardena2022}, is also thought to contribute to the observed variation in $\Av^{C}$.
To investigate this, we utilize the 3D dust maps from \citet[][hereafter \citetalias{Lallement2022}]{Lallement2022}\footnote{
Online G-tomo tool: \url{https://explore-platform.eu/sda/instances/d3ae1892-b30b-4281-9ef8-2d482aa69489_g-tomo}, 
using the cube v2 with a resolution of 10\,pc and a 5\,pc sampling.} and 
\citet[][hereafter \citetalias{Edenhofer2024}]{Edenhofer2024}\footnote{Data access via Zenodo, doi: 
\url{https://zenodo.org/records/10658339}.} to identify and count the number of the principal dust components along the sightline 
to each background star. For a given sightline, we identify overdensities in the dust profile using the 
{\it find\_peaks} function within the {\it scipy} package. 
To account for the resolution and uncertainties of the maps, we define a peak as a feature with a density exceeding 
0.3\,mag per kpc and a width of at least 3\,pc. The total number of components identified in front of a background star 
is denoted as $N_{\rm peaks}$. The upper panel of Fig. \ref{fig:peak1} presents an illustrative example for the 
sightline toward $(\ell,b)=(216.9^{\circ},{-}18.6^{\circ})$ in Orion A, with a background star at a distance of 936\,pc. 
Both dust maps reveal broadly consistent primary dust components along this sightline, although minor discrepancies 
exist due to their distinct reconstruction methodologies, underlying datasets, and modeling assumptions. 
For instance, while both maps detect two dust components between 400 and 500\,pc, two separate components between 700 and 
800\,pc resolved in the \citetalias{Lallement2022} map were treated as a single broad component in the \citetalias{Edenhofer2024} 
map, which additionally identified a component within 200\,pc. In general, the \citetalias{Lallement2022} and 
\citetalias{Edenhofer2024} maps yield consistent values for $N_{\rm peaks}$. As shown in the lower panel of Fig. \ref{fig:peak1}, 
the $N_{\rm peaks}$ values derived from the two maps agree to within $\pm$1 for 74\% of sightlines. Larger discrepancies typically 
arise when broad dust structures are resolved into a different number of narrower components by each map or from differences 
in the modeled density profiles at larger distances.

Although the investigated sightlines often traverse multiple clouds (see Fig. \ref{fig:peak2}), the velocity cut 
applied in our selection criteria limits the velocity separation between individual DIB components to be within 50\,$\kms$. 
This velocity difference corresponds to a maximum wavelength separation of approximately 1.4\,{\AA} between the centers 
of the DIB profiles originating in different clouds, which is smaller than the intrinsic Gaussian width of the $\lambda$8621 
profile \citep{hz2024a} and approximately twice of the spectral resolution of the GAIA RVS spectra.
Consequently, while the presence of multiple components will broaden the observed profile, the total 
$\Wgaia$ derived from the single Gaussian fit employed in \citet{Schultheis2023DR3} is expected to be only minimally affected.

Figure \ref{fig:peak2} displays the distribution of $N_{\rm peaks}$ for all investigated sightlines within each target 
cloud, as derived from the \citetalias{Edenhofer2024} map. Clouds with higher values of $N_{\rm peaks}$ tend to exhibit a larger 
$\Av^{C}$. For instance, in the Hercules and Mon R2 clouds, where $\Av^{C}\,{\sim}\,0.4{-}0.5$\,mag, $N_{\rm peaks}$ is
predominantly less than 4. In contrast, the Ophiuchus, Orion A, and California clouds, which are characterized by 
$\Av^{C}\,{>}\,2$\,mag, include a significant number of sightlines with $N_{\rm peaks}\,{>}\,5$. This observed contribution of 
$N_{\rm peaks}$ to the variation in $\Av^{C}$ can be explained by a model in which a high line-of-sight $\Av$ may result from 
the summation of many lower-density clumps rather than a single dense core. In such a scenario, UV radiation can penetrate the 
inter-clump medium more effectively, allowing DIB carriers to persist to a greater total column density than would be expected 
in a homogeneous cloud, 
resulting in a larger $\Av^{C}$. However, this trend is not universal, indicating that $\Av^{C}$ is not solely 
determined by the clumpy structure of the cloud. A notable exception is Orion B, which exhibits a greater number of peaks 
than Orion A, yet has a smaller $\Av^{C}$.

The projection of multiple structures along the sightline would also contribute to the scatter in the $\SEx$--$\SEy$ relations. 
A distinct trend within the $N_{\rm peaks}$ distribution is evident for the Ophiuchus, Chamaeleon, and California clouds 
(see Fig. \ref{fig:peak2}): for a given $\Av$, sightlines with larger $N_{\rm peaks}$ tend to exhibit a larger $\SEy$. 
Furthermore, when considering clouds with $n_p\,{=}\,2$, $\alpha_2$ increases across the sequence of clouds: Cepheus South, 
Taurus, Orion B, Ophiuchus, and California. This trend aligns with the increasing prevalence of sightlines with high $N_{\rm peaks}$ values observed in these same clouds. This finding is consistent with the suggestion by \citet{EL2019}, who proposed 
that a greater number of components along a sightline can flatten the observed trend, thereby yielding a larger value of 
$\alpha$, though in their investigated regions, Cepheus exhibited more complex structures than Orion and Taurus\footnote{Our 
target regions are much smaller than those in \citet{EL2019}. The Gaia RVS observations also probe shallower regions than
APOGEE.}.
Conversely, some clouds deviate from this pattern. Based on their relatively simple structures (predominantly 
$N_{\rm peaks}\,{\leqslant}\,3$), Canis Major and Chamaeleon were expected to have very small $\alpha_2$ values, 
corresponding to a steep trend. However, their measured $\alpha_2$ values are intermediate, falling between those 
of Taurus and Orion B.

Several clouds in our sample--Chamaeleon, Taurus, Perseus, Cepheus South, and Canis Major--exhibit regions where the slope is 
steeper than --1 ($\alpha\,{<}\,{-}1$). This indicates that $\Wgaia$ is actively decreasing with increasing $\Av$, a behavior 
that could be caused by the depletion of carrier molecules onto the surfaces of dust grains or by geometric effects, such as a decreasing effective path length through the carrier-rich region of the cloud. 
Perseus shows a particularly steep slope ($\alpha_2$) that subsequently 
transitions to a plateau ($\alpha_3\,{\approx}\,0$) for $\Av\,{\gtrsim}\,1.5$\,mag. A similar, though less well-sampled, flattening 
is observed in Mon R2. A zero slope in the $\SEy$--$\SEx$ plane corresponds to a linear correlation between $\Wgaia$ and $\Av$, a 
trend often seen over large distance ranges through the diffuse ISM \citep[e.g.,][]{Munari2008,hz2021a,hz2021b}. 
The structure along sightlines toward Perseus is highly uniform, with most characterized by $N_{\rm peaks}\,{=}\,2$.
Crucially, these two dust components do not indicate internal clumps within Perseus; one originates from the Perseus cloud 
itself, while the other is associated with the wall of the Local Bubble. Thus, the re-emergence of this linear trend at 
high $\Av$ cannot be attributed to cloud clumpiness.
At present, we observe this behavior only in Perseus and Mon R2 at $\Av\,{\sim}\,3$\,mag,
and these detections are based on a small number of sightlines that exhibit significant scatter.
Further observations probing more sightlines deeper into clouds are needed to confirm 
whether this re-increase in DIB strength is a genuine astrophysical effect, to assess 
its prevalence, and ultimately, to chart the evolution of DIBs in dense environments.

In summary, the observed behavior of the DIB\,$\lambda$8621 is broadly consistent with the predictions of the  
\citetalias{Sonnentrucker1997} PIE model, but the details are modulated by a combination of factors. 
The variation of the slope $\alpha$ both within a single cloud and between different clouds depends not only on 
the UV field and electron density but also critically on the geometry of the clouds, including their large-scale 
density profiles and small-scale clumpy substructures. 
Future investigations focusing on nearby, structurally simple clouds--utilizing high-resolution archival spectra 
of early-type stars capable of resolving individual DIB components and probing denser regions--will be invaluable 
for disentangling these effects and further constraining the properties of DIB carriers.
Ultimately, a comprehensive understanding of these relationships could allow 
the variation of $\Wdib$ with $\Av$ to be used as a powerful tracer of both the UV radiation field and cloud structure 
in the diffuse and translucent ISM.

\section{Summary and conclusions} \label{sect:conclusion}

With measurements of DIB\,$\lambda$8621 from Gaia DR3 and $\Av$ derived from GSP-Spec stellar parameters, we have investigated 
the variation of the normalized DIB strength, $\SEy$, as a function of extinction, $\SEx$, toward 12 nearby molecular clouds, 
spanning diffuse to translucent regions ($\Av \sim 0.2{-}3.5$\,mag). Our analysis reveals significant diversity in the DIB behavior, 
both from one cloud to another and from the outer to the inner regions of a single cloud. We modeled these trends using a piecewise 
linear model (PLM), fitting the average slope ($\alpha$) of the $\SEy$--$\SEx$ relation over different extinction ranges and selecting 
the optimal number of segments for each cloud.

For most investigated clouds, the relationship between $\SEy$ and $\SEx$ shows an initial decline with an average slope between 0 and 
--1. This trend reflects a balance between the accumulation of precursor gas and the attenuation of the interstellar UV field with 
increasing cloud depth. The observed diversity and variation of slopes can be attributed to several key environmental factors:

\begin{enumerate}
    \item The ambient UV radiation field dictates the photoionization equilibrium of the DIB carrier. As dust accumulates deeper 
    inside a cloud, the UV field is attenuated, causing the carrier abundance to drop. When the UV field is effectively extinguished, 
    $\Wgaia$ becomes nearly constant, and the slope $\alpha$ approaches --1. The critical extinction ($\Av^C$) at which this occurs 
    varies widely among clouds, from $\sim$0.4\,mag (e.g., Hercules) to $>$3\,mag (e.g., Orion A).
    \item The cloud geometry--including the gas and dust density profiles and the presence of clumpy substructures--influences the 
    rate of UV attenuation, the effective line-of-sight path length, and the relationship between total column density and local 
    volume density for DIB carriers and dust grains. A longer path length or projection effects from multiple clumps can increase 
    the total carrier column density (and thus $\Wgaia$), resulting in a shallower slope. 
    Clumpiness also contributes to the variation of $\Av^C$ and the scatter in the $\SEx$--$\SEy$ relation.
    \item The electron density ($N_e$), which decreases significantly in the transition to translucent gas, affects the recombination 
    rate of the ionized carrier. A lower $N_e$ reduces recombination efficiency, allowing the carrier to survive to greater depths 
    and thus leading to a shallower slope.
    \item The depletion of carrier molecules onto dust grain surfaces is a potential explanation for clouds exhibiting slopes steeper 
    than --1 ($\alpha\,{<}\,{-}1$), such as Taurus and Cepheus South, where $\Wgaia$ actively decreases with $\Av$.
\end{enumerate}

Taurus is the only cloud in our sample that exhibits the initial rising trend of ${\rm log_{10}}(\Wgaia/\EBV)$ at low extinction 
($\EBV\,{<}\,0.2$\,mag), as predicted by the PIE model for an ionized carrier. A linear fit to this trend yields a slope of 
$\alpha_T\,{=}\,3.24_{{-}1.10}^{+1.30}$. Assuming $\Rv\,{=}\,3.1$, this slope corresponds to an ionization potential of 
$E_{\rm IP}\,{=}\,12.40^{+1.90}_{-2.29}$\,eV
for the DIB\,$\lambda$8621 carrier. This value is consistent with the secondary ionization energies of large molecules like PAHs 
and fullerenes. This finding supports the interpretation that the $\lambda$8621 carrier is likely a cation, similar to the proposed 
nature of the optical DIBs at $\lambda$5780, $\lambda$5797, and $\lambda$6614 in \citetalias{Sonnentrucker1997}.

By characterizing the $\EBV$ at which ${\rm log_{10}}(\Wgaia/\EBV)$ peaks in Taurus and comparing this with the results for other 
DIBs from \citetalias{Sonnentrucker1997}, we propose a spatial sequence of DIBs from the outer to the inner regions of a cloud: 
$\lambda5780 \rightarrow \lambda8621 \rightarrow \lambda6614 \rightarrow \lambda5797 \rightarrow \lambda6379$. This sequence aligns 
with the decreasing of their normalized $\Wdib$ reported in \citet{Fan2019}.

Together, these results demonstrate that the relationship between DIB\,$\lambda$8621 and dust extinction provides powerful 
constraints on the properties of its carrier and serves as a sensitive probe of cloud structure and local physical 
conditions. Furthermore, comparing the relative behaviors of different DIBs across opacity regimes emerges as a promising avenue 
for unraveling their shared and distinct characteristics.

\begin{acknowledgements}
We thank the referee for a comprehensive review and insightful feedback, which included constructive 
suggestions that helped to reinforce our results.
H.Z. acknowledges financial support by the Chilean Government--ESO Joint Committee (Comit\'{e} Mixto ESO--Chile) through project 
No. annlang23003-es-cl.
L.Li thanks the support of Natural Science Foundation of China No. 12303026 and the Young Data Scientist Project of the National 
Astronomical Data Center. 
This work has made use of data from the European Space Agency (ESA) mission Gaia (\url{https://www.cosmos.esa.int/gaia}), 
processed by the Gaia Data Processing and Analysis Consortium (DPAC, \url{https://www.cosmos.esa.int/web/gaia/dpac/consortium}). 
Funding for the DPAC has been provided by national institutions, in particular, the institutions participating in the Gaia Multilateral 
Agreement.
\end{acknowledgements}

\bibliographystyle{aa}
\bibliography{reference}

\appendix

\section{Individual target cloud regions} \label{appsec:mcs}

\begin{figure*}[!htp]
  \centering
  \includegraphics[width=16cm]{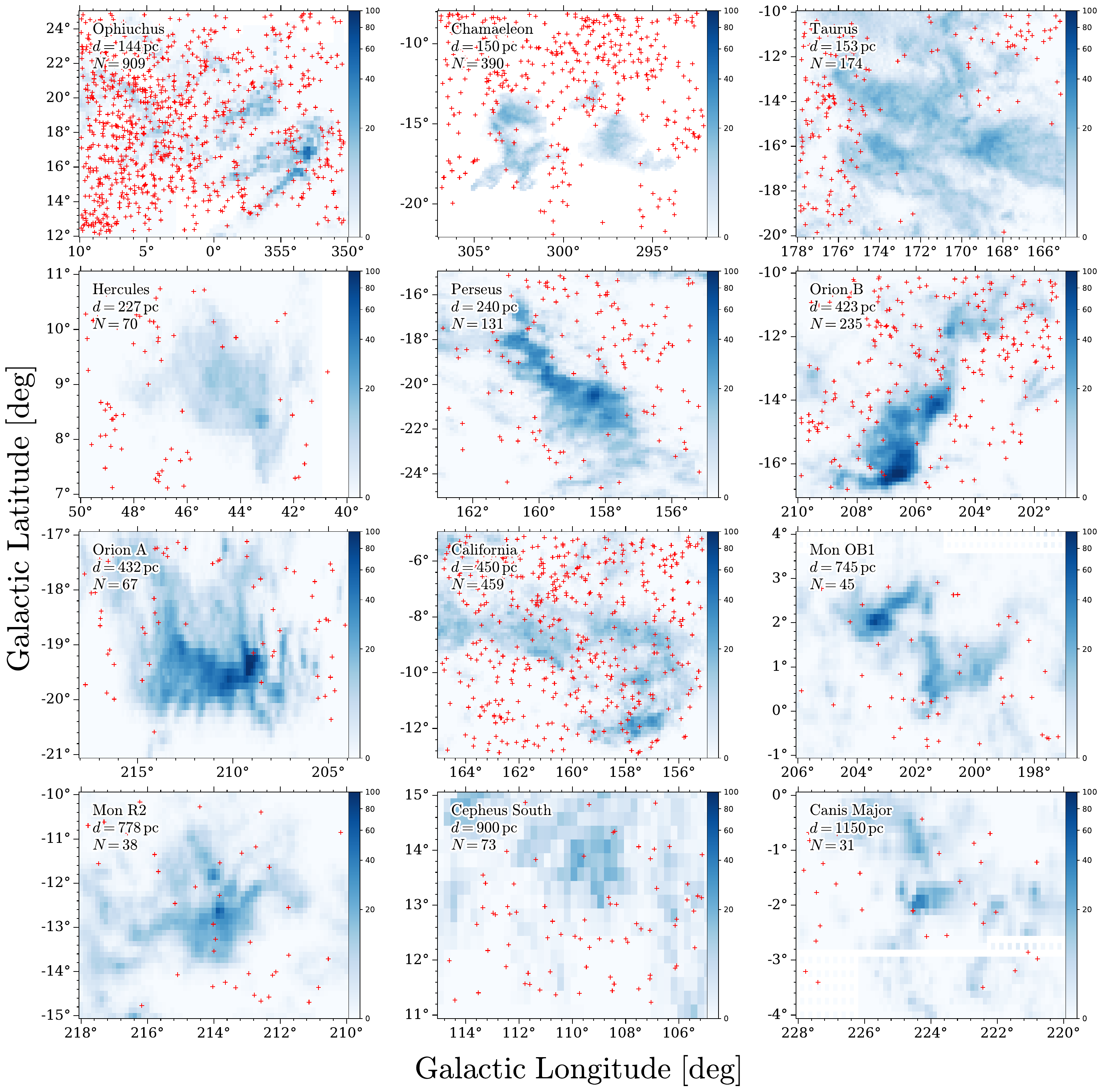}
  \caption{Distribution of selected sightlines (red crosses) for each cloud, overlaid on the $^{12}\rm CO$ ($J\,{=}\,1{-}0$) 
  integrated intensity map from \citet{Dame2001}. The cloud name, distance, and number of selected sightlines are labeled in each 
  panel.}
  \label{fig:mcs}
\end{figure*}

Figure \ref{fig:mcs} displays the spatial distribution of the selected sightlines for each of the 12 target clouds. The sightlines, 
for which reliable $\Wgaia$ and $\Av$ measurements were obtained, are shown as red crosses. These are overlaid on the $^{12}\rm CO$ 
($J\,{=}\,1{-}0$) integrated intensity maps from \citet{Dame2001}. Due to the limiting magnitude of Gaia RVS spectra and our data 
quality control, the majority of sightlines are located in the diffuse to translucent regions surrounding the denser parts of the 
clouds.

\section{Complete PLM Fitting Results for Each Target Cloud} \label{appsec:plm}

This appendix presents the complete set of fitting results for the piecewise linear models (PLMs) applied to each cloud. Figure 
\ref{fig:mc-plm-all} shows the results for PLMs with 1--4 linear segments. In each panel, the model selected as the optimal 
representation (indicated by a green check mark) is the one presented in Fig. \ref{fig:skin-effect} in the main text. The full set 
of fitting parameters (intercept $C$, slopes $\alpha$, and knots $k$) for all test models are provided in Table \ref{tab:plm-full}. 
For convenience, the parameters corresponding to only the optimal models are collected in Table \ref{tab:plm}. We note that for 
Cepheus South, Fig. \ref{fig:mc-plm-all} displays the fits both with and without outlier clipping, though Table \ref{tab:plm-full} 
lists only the parameters derived from the fit with clipping.

\begin{table*}
\centering
\renewcommand{\arraystretch}{1.45}
\tiny
\caption{Fitting results of PLM with $n_p\,{=}\,1{-}4$ for each target cloud. }
\label{tab:plm-full}
\begin{tabular}{l c *{8}{r@{}l}} 
\toprule
Name & $n_p$ & \multicolumn{2}{c}{$C$} & \multicolumn{2}{c}{$\alpha_1$} & \multicolumn{2}{c}{$\alpha_2$} & 
\multicolumn{2}{c}{$\alpha_3$} & \multicolumn{2}{c}{$\alpha_4$} & \multicolumn{2}{c}{$k_1$} & \multicolumn{2}{c}{$k_2$} & \multicolumn{2}{c}{$k_3$} \\ 
\hline 
Ophiuchus & 1 &  $2.16$ & {$_{-0.01}^{+0.01}$} & $-0.80$ & {$_{-0.01}^{+0.01}$} & & & & & & & & & & & & \\ 
          & 2 &  $2.21$ & {$_{-0.01}^{+0.01}$} & $-0.49$ & {$_{-0.02}^{+0.02}$} & $-0.90$ & {$_{-0.05}^{+0.05}$} & & & & & 
                 $-0.07$ & {$_{-0.01}^{+0.01}$} & & & & \\ 
          & 3 &  $2.00$ & {$_{-0.01}^{+0.01}$} & $-1.45$ & {$_{-0.02}^{+0.02}$} & $13.47$ & {$_{-0.17}^{+0.08}$} & 
                 $-1.19$ & {$_{-0.23}^{+0.23}$} & & & $-0.02$ & {$_{-0.01}^{+0.01}$} & $-0.00$ & {$_{-0.01}^{+0.01}$} & & \\ 
          & 4 &  $2.17$ & {$_{-0.01}^{+0.01}$} & $-0.64$ & {$_{-0.01}^{+0.01}$} & $-14.83$ & {$_{-0.24}^{+0.50}$} & $-0.14$ & {$_{-0.72}^{+0.73}$} & $-2.99$ & {$_{-1.10}^{+1.10}$} & $0.20$ & {$_{-0.01}^{+0.01}$}  & $0.21$ & {$_{-0.01}^{+0.01}$}  & $0.42$ & {$_{-0.01}^{+0.01}$}  \\ 
\hline
Chamaeleon & 1 & $2.03$ & {$_{-0.01}^{+0.01}$} & $-0.99$ & {$_{-0.02}^{+0.02}$} & & & & & & & & & & & & \\ 
           & 2 & $2.18$ & {$_{-0.02}^{+0.02}$} & $-0.57$ & {$_{-0.06}^{+0.05}$} & $-1.13$ & {$_{-0.12}^{+0.12}$} & & & & & $-0.31$ & {$_{-0.01}^{+0.01}$} & & & & \\ 
           & 3 & $2.08$ & {$_{-0.01}^{+0.01}$} & $-0.81$ & {$_{-0.03}^{+0.03}$} & $-2.04$ & {$_{-0.16}^{+0.15}$} & $12.49$ & {$_{-0.84}^{+0.49}$} & & & $-0.11$ & {$_{-0.02}^{+0.01}$} & $0.19$ & {$_{-0.01}^{+0.01}$}  & & \\ 
           & 4 & $2.00$ & {$_{-0.01}^{+0.01}$} & $-0.97$ & {$_{-0.02}^{+0.02}$} & $12.14$ & {$_{-0.92}^{+0.51}$} & $-2.21$ & {$_{-1.40}^{+1.43}$} & $12.36$ & {$_{-2.07}^{+1.75}$} & $-0.21$ & {$_{-0.01}^{+0.01}$} & $-0.20$ & {$_{-0.01}^{+0.01}$} & $0.18$ & {$_{-0.01}^{+0.01}$}  \\ 
\hline
Taurus	   & 1 & $2.10$ & {$_{-0.01}^{+0.01}$} & $-1.13$ & {$_{-0.02}^{+0.02}$} & & & & & & & & & & & & \\ 
           & 2 & $2.14$ & {$_{-0.01}^{+0.01}$} & $-0.59$ & {$_{-0.06}^{+0.06}$} & $-1.30$ & {$_{-0.13}^{+0.13}$} & & & & & $-0.01$ &                 {$_{-0.01}^{+0.01}$} & & & & \\ 
           & 3 & $2.00$ & {$_{-0.01}^{+0.01}$} & $-1.29$ & {$_{-0.05}^{+0.05}$} & $12.48$ & {$_{-1.94}^{+0.86}$} & $-1.44$ &                         {$_{-2.76}^{+2.73}$} & & & $-0.03$ & {$_{-0.01}^{+0.01}$} & $-0.02$ & {$_{-0.01}^{+0.01}$} & & \\ 
           & 4 & $2.15$ & {$_{-0.01}^{+0.01}$} & $-0.45$ & {$_{-0.05}^{+0.05}$} & $-2.55$ & {$_{-0.23}^{+0.24}$} & $12.22$ & {$_{-0.60}^{+0.41}$} & $-2.10$ & {$_{-0.96}^{+0.85}$} & $0.04$ & {$_{-0.01}^{+0.01}$} & $0.23$ & {$_{-0.01}^{+0.01}$} & $0.26$ & {$_{-0.01}^{+0.01}$}  \\
\hline
Hercules   & 1 & $2.03$ & {$_{-0.01}^{+0.01}$} & $-1.06$ & {$_{-0.05}^{+0.05}$} & & & & & & & & & & & & \\ 
           & 2 & $4.32$ & {$_{-0.41}^{+0.24}$} & $4.27$ & {$_{-0.91}^{+0.54}$} & $-1.24$ & {$_{-1.45}^{+1.43}$} & & & & & $-0.42$ &                  {$_{-0.01}^{+0.01}$} & & & & \\ 
           & 3 & $2.00$ & {$_{-0.01}^{+0.01}$} & $-1.08$ & {$_{-0.03}^{+0.03}$} & $7.06$ & {$_{-1.51}^{+1.56}$} & $-6.72$ &                          {$_{-2.39}^{+3.11}$} & & & $-0.13$ & {$_{-0.01}^{+0.01}$} & $-0.09$ & {$_{-0.01}^{+0.01}$} & & \\ 
           & 4 & $4.23$ & {$_{-0.46}^{+0.29}$} & $4.10$ & {$_{-1.01}^{+0.65}$} & $-1.85$ & {$_{-1.67}^{+1.65}$} & $6.85$ & {$_{-3.23}^{+3.23}$} & $-6.60$ & {$_{-4.31}^{+4.81}$} & $-0.40$ & {$_{-0.01}^{+0.01}$} & $-0.15$ & {$_{-0.01}^{+0.01}$} & $-0.09$ & {$_{-0.01}^{+0.01}$}  \\
\hline
Perseus	   & 1 & $2.12$ & {$_{-0.01}^{+0.01}$} & $-0.87$ & {$_{-0.01}^{+0.01}$} & & & & & & & & & & & & \\ 
           & 2 & $2.11$ & {$_{-0.01}^{+0.01}$} & $-0.98$ & {$_{-0.03}^{+0.02}$} & $-0.27$ & {$_{-0.15}^{+0.15}$} & & & & & $0.26$ &                  {$_{-0.02}^{+0.02}$} & & & & \\ 
           & 3 & $2.18$ & {$_{-0.01}^{+0.01}$} & $-0.53$ & {$_{-0.04}^{+0.04}$} & $-2.40$ & {$_{-0.25}^{+0.23}$} & $-0.06$ &                         {$_{-0.46}^{+0.45}$} & & & $0.03$ & {$_{-0.01}^{+0.01}$} & $0.16$ & {$_{-0.01}^{+0.01}$} & & \\ 
           & 4 & $2.18$ & {$_{-0.01}^{+0.01}$} & $-0.54$ & {$_{-0.04}^{+0.04}$} & $-2.06$ & {$_{-0.18}^{+0.15}$} & $0.51$ & {$_{-0.40}^{+0.38}$} & $-3.61$ & {$_{-6.78}^{+3.22}$} & $0.02$ & {$_{-0.01}^{+0.01}$} & $0.21$ & {$_{-0.01}^{+0.01}$} & $0.46$ & {$_{-0.06}^{+0.02}$}  \\
\hline
Orion B	   & 1 & $2.20$ & {$_{-0.01}^{+0.01}$} & $-0.71$ & {$_{-0.01}^{+0.01}$} & & & & & & & & & & & & \\ 
           & 2 & $2.23$ & {$_{-0.01}^{+0.02}$} & $-0.23$ & {$_{-0.05}^{+0.11}$} & $-0.95$ & {$_{-0.12}^{+0.17}$} & & & & & $0.03$ &                  {$_{-0.04}^{+0.01}$} & & & & \\ 
           & 3 & $2.26$ & {$_{-0.01}^{+0.01}$} & $-0.01$ & {$_{-0.05}^{+0.04}$} & $-14.33$ & {$_{-0.54}^{+1.04}$} & $-0.62$ &                        {$_{-1.51}^{+1.51}$} & & & $0.07$ & {$_{-0.01}^{+0.01}$} & $0.08$ & {$_{-0.01}^{+0.01}$} & & \\ 
           & 4 & $2.17$ & {$_{-0.01}^{+0.01}$} & $-0.78$ & {$_{-0.01}^{+0.01}$} & $7.25$ & {$_{-0.01}^{+0.01}$} & $-7.74$ & {$_{-0.01}^{+0.01}$} & $-0.87$ & {$_{-0.01}^{+0.01}$} & $0.12$ & {$_{-0.01}^{+0.01}$} & $0.16$ & {$_{-0.01}^{+0.01}$} & $0.20$ & {$_{-0.01}^{+0.01}$}  \\
\hline
Orion A	   & 1 & $2.07$ & {$_{-0.01}^{+0.01}$} & $-0.74$ & {$_{-0.02}^{+0.02}$} & & & & & & & & & & & & \\ 
           & 2 & $3.78$ & {$_{-1.72}^{+0.53}$} & $3.23$ & {$_{-4.04}^{+1.10}$} & $-0.77$ & {$_{-5.16}^{+5.29}$} & & & & & $-0.43$ &                  {$_{-0.01}^{+0.37}$} & & & & \\ 
           & 3 & $4.20$ & {$_{-0.39}^{+0.29}$} & $4.14$ & {$_{-0.90}^{+0.61}$} & $-3.77$ & {$_{-2.46}^{+1.93}$} & $-0.55$ &                          {$_{-3.57}^{+3.18}$} & & & $-0.39$ & {$_{-0.01}^{+0.01}$} & $-0.30$ & {$_{-0.03}^{+0.03}$} & & \\ 
           & 4 & $2.07$ & {$_{-0.02}^{+0.01}$} & $-0.78$ & {$_{-0.13}^{+0.05}$} & $0.81$ & {$_{-1.28}^{+4.03}$} & $-7.72$ & {$_{-5.25}^{+8.65}$} & $-13.09$ & {$_{-11.65}^{+19.67}$} & $0.32$ & {$_{-0.28}^{+0.04}$} & $0.40$ & {$_{-0.02}^{+0.01}$} & $0.42$ & {$_{-0.01}^{+0.01}$}  \\
\hline
California & 1 & $2.18$ & {$_{-0.01}^{+0.01}$} & $-0.68$ & {$_{-0.01}^{+0.01}$} & & & & & & & & & & & & \\ 
           & 2 & $2.18$ & {$_{-0.01}^{+0.01}$} & $-0.46$ & {$_{-0.02}^{+0.02}$} & $-0.87$ & {$_{-0.05}^{+0.05}$} & & & & & $0.14$ &                  {$_{-0.01}^{+0.01}$} & & & & \\ 
           & 3 & $2.03$ & {$_{-0.01}^{+0.01}$} & $-1.49$ & {$_{-0.06}^{+0.06}$} & $13.36$ & {$_{-0.28}^{+0.17}$} & $-1.03$ &                         {$_{-0.41}^{+0.40}$} & & & $0.02$ & {$_{-0.01}^{+0.01}$} & $0.04$ & {$_{-0.01}^{+0.01}$} & & \\ 
           & 4 & $2.35$ & {$_{-0.05}^{+0.05}$} & $0.05$ & {$_{-0.17}^{+0.21}$} & $-3.74$ & {$_{-0.54}^{+0.51}$} & $11.08$ & {$_{-0.88}^{+0.64}$} & $-0.88$ & {$_{-1.24}^{+1.23}$} & $-0.15$ & {$_{-0.01}^{+0.01}$} & $-0.05$ & {$_{-0.01}^{+0.01}$} & $-0.02$ & {$_{-0.01}^{+0.01}$}  \\
\hline
Mon OB1	   & 1 & $2.23$ & {$_{-0.01}^{+0.01}$} & $-0.39$ & {$_{-0.04}^{+0.03}$} & & & & & & & & & & & & \\ 
           & 2 & $2.18$ & {$_{-0.01}^{+0.01}$} & $-0.86$ & {$_{-0.05}^{+0.05}$} & $13.65$ & {$_{-0.70}^{+0.40}$} & & & & & $0.31$ &                  {$_{-0.01}^{+0.01}$} & & & & \\ 
           & 3 & $2.19$ & {$_{-0.01}^{+0.01}$} & $-0.77$ & {$_{-0.05}^{+0.05}$} & $7.54$ & {$_{-1.44}^{+1.34}$} & $-6.50$ &                          {$_{-2.14}^{+2.47}$} & & & $0.25$ & {$_{-0.01}^{+0.01}$} & $0.29$ & {$_{-0.01}^{+0.01}$} & & \\ 
           & 4 & $3.42$ & {$_{-0.32}^{+0.23}$} & $3.70$ & {$_{-1.08}^{+0.81}$} & $-0.89$ & {$_{-1.87}^{+1.90}$} & $6.99$ & {$_{-3.40}^{+3.44}$} & $-6.95$ & {$_{-4.20}^{+5.06}$} & $-0.26$ & {$_{-0.01}^{+0.01}$} & $0.24$ & {$_{-0.01}^{+0.01}$} & $0.29$ & {$_{-0.01}^{+0.01}$}  \\
\hline
Mon R2	   & 1 & $2.16$ & {$_{-0.01}^{+0.01}$} & $-0.82$ & {$_{-0.06}^{+0.05}$} & & & & & & & & & & & & \\ 
           & 2 & $2.16$ & {$_{-0.01}^{+0.01}$} & $-0.96$ & {$_{-0.09}^{+0.09}$} & $0.09$ & {$_{-0.47}^{+0.96}$} & & & & & $0.25$ &                   {$_{-0.05}^{+0.14}$} & & & & \\ 
           & 3 & $2.16$ & {$_{-0.01}^{+0.01}$} & $-0.92$ & {$_{-0.10}^{+0.08}$} & $1.65$ & {$_{-2.06}^{+7.60}$} & $-0.61$ &                          {$_{-9.91}^{+17.35}$} & & & $0.30$ & {$_{-0.07}^{+0.14}$} & $0.44$ & {$_{-0.10}^{+0.02}$} & & \\ 
           & 4 & $2.16$ & {$_{-0.01}^{+0.01}$} & $-0.90$ & {$_{-0.10}^{+0.11}$} & $-5.71$ & {$_{-4.36}^{+7.57}$} & $3.53$ & {$_{-17.70}^{+11.48}$} & $1.46$ & {$_{-21.69}^{+18.15}$} & $0.14$ & {$_{-0.03}^{+0.22}$} & $0.17$ & {$_{-0.03}^{+0.28}$} & $0.24$ & {$_{-0.06}^{+0.23}$}  \\
\hline
Cepheus South & 1 & $2.32$ & {$_{-0.01}^{+0.01}$} & $-0.80$ & {$_{-0.04}^{+0.03}$} & & & & & & & & & & & & \\ 
              & 2 & $2.31$ & {$_{-0.01}^{+0.01}$} & $-0.57$ & {$_{-0.07}^{+0.07}$} & $-1.46$ & {$_{-0.38}^{+0.27}$} & & & & & $0.22$ &               {$_{-0.04}^{+0.04}$} & & & & \\ 
              & 3 & $2.29$ & {$_{-0.01}^{+0.01}$} & $-0.91$ & {$_{-0.12}^{+0.11}$} & $7.83$ & {$_{-3.16}^{+3.89}$} & $-1.47$ &                       {$_{-6.94}^{+6.90}$} & & & $0.10$ & {$_{-0.01}^{+0.01}$} & $0.12$ & {$_{-0.01}^{+0.01}$} & & \\ 
              & 4 & $2.29$ & {$_{-0.01}^{+0.01}$} & $-0.89$ & {$_{-0.11}^{+0.12}$} & $8.58$ & {$_{-3.08}^{+2.86}$} & $-1.49$ & {$_{-5.79}^{+5.82}$} & $-2.16$ & {$_{-12.05}^{+9.72}$} & $0.10$ & {$_{-0.01}^{+0.01}$} & $0.12$ & {$_{-0.01}^{+0.01}$} & $0.45$ & {$_{-0.05}^{+0.01}$}  \\
\hline
Canis Major	  & 1 & $2.32$ & {$_{-0.01}^{+0.01}$} & $-1.03$ & {$_{-0.03}^{+0.03}$} & & & & & & & & & & & & \\ 
              & 2 & $2.42$ & {$_{-0.01}^{+0.01}$} & $-0.30$ & {$_{-0.07}^{+0.07}$} & $-1.20$ & {$_{-0.16}^{+0.16}$} & & & & & $-0.08$ &              {$_{-0.01}^{+0.01}$} & & & & \\ 
              & 3 & $2.43$ & {$_{-0.01}^{+0.01}$} & $-0.29$ & {$_{-0.07}^{+0.07}$} & $-1.77$ & {$_{-0.77}^{+0.50}$} & $-0.87$ &                      {$_{-1.25}^{+1.24}$} & & & $-0.06$ & {$_{-0.02}^{+0.02}$} & $0.05$ & {$_{-0.04}^{+0.10}$} & & \\ 
              & 4 & $3.28$ & {$_{-0.37}^{+0.53}$} & $1.92$ & {$_{-0.99}^{+1.38}$} & $-1.05$ & {$_{-2.37}^{+2.38}$} & $-6.37$ & {$_{-5.55}^{+4.62}$} & $3.95$ & {$_{-8.97}^{+7.61}$} & $-0.32$ & {$_{-0.02}^{+0.03}$} & $0.30$ & {$_{-0.02}^{+0.02}$} & $0.35$ & {$_{-0.02}^{+0.02}$}  \\
\bottomrule
\end{tabular}
\tablefoot{\tiny
Columns: $n_p$: Number of component for the PLM; $C$: intercept;
$\alpha_1-\alpha_4$: fitted and calculated slopes of each linear segment in PLM ($ \alpha_{i+1} = \alpha_1 + \sum_{j=1}^{i} \Delta \alpha_j $
for $i>1$); $k_1-k_4$: fitted knot position for each PLM.}
\end{table*}

\begin{figure*}[!htp]
  \centering
  \includegraphics[width=16cm]{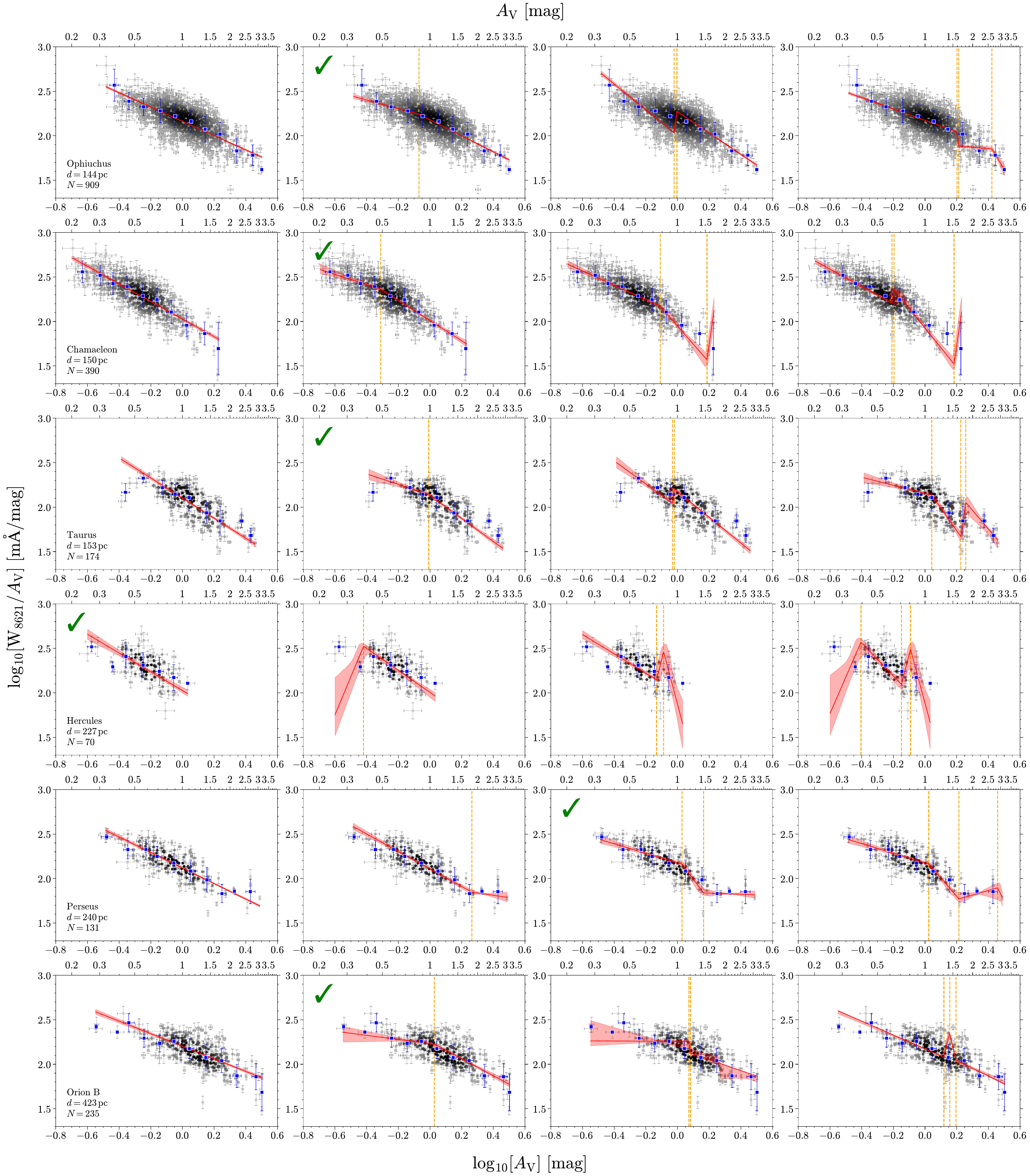}
  \caption{Fitting results of the piecewise linear model (PLM) with $n_p\,{=}\,1{-}4$ for each target cloud. The fit models 
  are shown as red lines with shaded confidence intervals, overlaid on the individual measurements. The median trend of $\SEy$ 
  as a function of $\SEx$ is shown by the blue squares. The locations of the knots ($k$) for each PLM are indicated by dashed 
  orange lines. The model selected as optimal for each cloud is marked with a green check mark. The cloud name, distance, and 
  number of sightlines used in the fit are also marked for each cloud.}
  \label{fig:mc-plm-all}
\end{figure*}\addtocounter{figure}{-1}

\begin{figure*}[!htp]
  \centering
  \includegraphics[width=16cm]{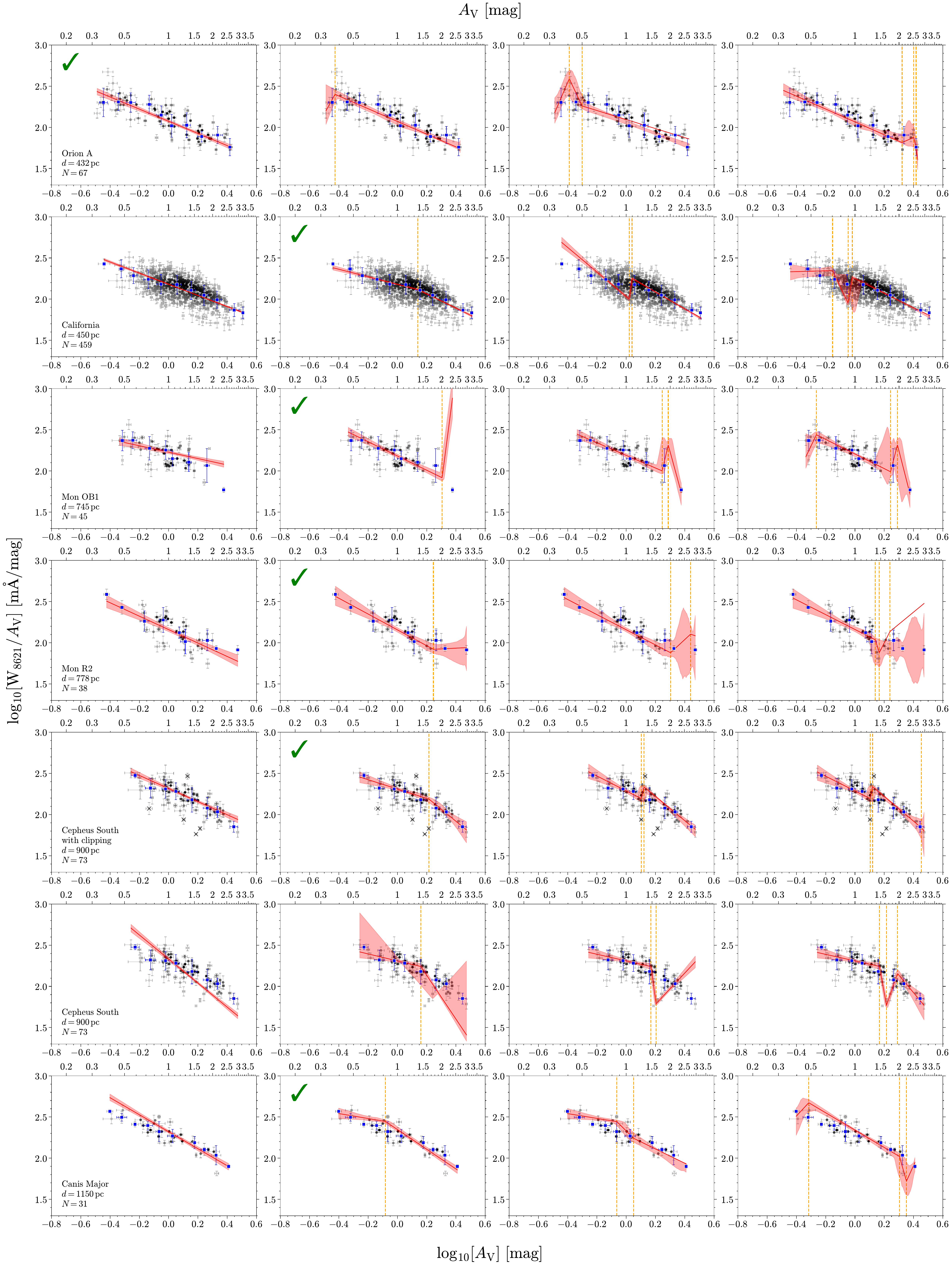}
  \caption{--continued.}
\end{figure*}

\clearpage

\end{CJK*}

\end{document}